\documentclass[12pt]{article}
\usepackage{amssymb}
\usepackage{amsfonts}
\usepackage{graphicx}
\usepackage{lscape}
\usepackage{amsmath}
\usepackage[onehalfspacing]{setspace}
\usepackage[letterpaper]{geometry}
\usepackage[sc]{mathpazo}
\usepackage{amsmath}
\usepackage{caption}
\usepackage{printlen}
\usepackage{blindtext}
\usepackage{graphicx}
\usepackage{makecell}
\usepackage{rotating}
\usepackage{bm}
\usepackage{tabularx}
\usepackage{bigints}
\usepackage{color}
\usepackage{xcolor}
\usepackage{graphicx}
\usepackage{bbm}
\usepackage{enumitem}
\usepackage{caption}
\usepackage{subfig}
\usepackage{dsfont}
\usepackage[nodayofweek]{datetime}

\usepackage{hyperref}  
\hypersetup{colorlinks,linkcolor={blue},citecolor={blue},urlcolor={blue}}

\usepackage{cleveref}
\usepackage{mathtools, nccmath, relsize}
\usepackage[natbibapa]{apacite}
\setcitestyle{authoryear,open={(},close={)}}
\usepackage{url}
\usepackage{titlesec}
\usepackage[title]{appendix}
\usepackage{graphicx,dblfloatfix}
\usepackage{float} 
\usepackage[english]{babel}
\catcode`"=13
\def"-{\allowhyphens-}

\usepackage{datetime}
\usdate

\usepackage[flushleft]{threeparttable}

\usepackage{pgf, tikz, tkz-tab}
\usetikzlibrary{calc, intersections, through, backgrounds}
\usetikzlibrary{patterns}
\usetikzlibrary{decorations.pathreplacing}
\usepackage{tabu}
\usepackage{rotating, booktabs,caption}
\usepackage{dcolumn}
\usepackage{color}               
\usepackage{colortbl} 

\usepackage{multirow}

\usepackage[utf8]{inputenc}
\usepackage[T1]{fontenc}
\usepackage{tgtermes}
\usepackage{mathptmx}

\usepackage{textcomp}
\usepackage{gensymb}

\usepackage{adjustbox}

\newtheorem{theorem}{Theorem}

\newtheorem{assumption}{Assumption}

\newtheorem{claim}{Claim}

\newtheorem{definition}{Definition}
\newtheorem{example}{Example}

\newtheorem{lemma}{Lemma}

\newtheorem{proposition}{Proposition}

\newenvironment{proof}[1][Proof]{\textbf{#1.} }{\ \rule{0.6em}{0.6em}}
\expandafter\def\expandafter\normalsize\expandafter{%
    \normalsize
    \setlength\abovedisplayskip{6pt}
    \setlength\belowdisplayskip{6pt}
    \setlength\abovedisplayshortskip{6pt}
    \setlength\belowdisplayshortskip{6pt}
}

\titlespacing*{\section}
{0pt}{0.5em}{0.5em}
\titlespacing*{\subsection}
{0pt}{0.5em}{0.5em}

\usepackage[flushleft]{threeparttable}
\usepackage{scalerel,stackengine}
\stackMath

\newcommand\reallywidehat[1]{%
\savestack{\tmpbox}{\stretchto{%
  \scaleto{%
    \scalerel*[\widthof{\ensuremath{#1}}]{\kern-.6pt\bigwedge\kern-.6pt}%
    {\rule[-\textheight/2]{1ex}{\textheight}}
  }{\textheight}%
}{0.5ex}}%
\stackon[1pt]{#1}{\tmpbox}%
}
\title{Recurring Auctions with Costly Entry: \\
Theory and Evidence}
\author{Shanglyu Deng, Qiyao Zhou\thanks{Deng: Department of Economics, University of Macau (email: \href{mailto:sdeng@um.edu.mo}{sdeng@um.edu.mo}); Zhou: Department of Economics, Oklahoma State University (email: \href{mailto:qzhou@okstate.edu}{qiyaoz@umd.edu}). 
Special thanks are due to our advisors Lawrence M. Ausubel and Ginger Zhe Jin for their guidance and support.  
We thank Arie Beresteanu, Maureen Cropper, Kate Ho, Charles Hodgson, Andrew Sweeting, Daniel Vincent, Chenyu Yang, and seminar/conference participants at the University of Maryland, Shanghai Jiao Tong University, 2024 IIOC Annual Conference, 2023 EARIE Annual Conference, 2023 Greater Bay Area Market Design Workshop, and 2023 Microeconomic Theory Workshop at Renmin University for helpful discussions, suggestions, and comments. Deng thanks the National Natural Science Foundation of China (No.\thinspace 72403270), the Start-up Research Grant (SRG2023-00036-FSS) of the University of Macau, and the 2024 seed grant from the Asia-Pacific Academy of Economics and Management (APAEM/SG/0002/2024) for financial support. Zhou gratefully acknowledges financial support from the Allan Gruchy fellowship. Any errors are our own.}}
\date{\today}

\begin{document}
\maketitle
\thispagestyle{empty}


\begin{abstract}
\begin{singlespace}
\noindent 
Recurring auctions are ubiquitous for selling durable assets such as artwork and homes, with follow-up auctions held for unsold items. We investigate such auctions theoretically and empirically. Theoretical analysis demonstrates that recurring auctions outperform single-round auctions when buyers face entry costs, enhancing efficiency and revenue due to sorted entry of potential buyers. Optimal reserve price sequences are characterized. Empirical findings from home foreclosure auctions in China reveal significant annual gains in efficiency (3.40 billion USD, 16.60\%) and revenue (2.97 billion USD, 15.92\%) using recurring auctions compared with single-round auctions. Implementing optimal reserve prices can further improve efficiency (3.35\%) and revenue (3.06\%).

\end{singlespace}

\begin{description}
\item[Keywords:] recurring auctions, auction design, sorting, entry.

\item[JEL Classification Codes:]  D44, D82, R31.
\end{description}
\end{abstract}

\clearpage
\pagenumbering{arabic}

\section{Introduction}
Auctions for durable assets such as houses and artwork are commonly recurring: A subsequent auction is often held if the initial one fails to sell the item. Despite the prevalence of recurring auctions, efforts to understand their existence and equilibrium properties have been scarce. This paper fills the gap by providing both a theoretical model and an empirical analysis of recurring auctions. We show that recurring auctions can outperform single-round mechanisms in terms of efficiency and revenue, and evaluate the related gains in an empirical setting. The results aid in explaining the ubiquity of such mechanisms. Our analyses also shed light on how to fine-tune recurring auctions to further promote efficiency and revenue.

Costly entry, which is usually the case in reality,\footnote{For example, it may take time and effort to qualify as a bidder and prepare a bid. In particular, a deposit is often required in auctions for valuable assets, which imposes a liquidity cost on the bidder.} underlies the dominance of recurring auctions over single-round ones. We illustrate this point by setting up an auction model with independent private values (IPV) and costly entry. In the model, a seller has one item for sale and there is a fixed pool of potential buyers whose values are independently drawn from the same distribution.
The seller holds auctions in a recurring manner: If an auction fails, either because no bidder shows up or because the reserve price is not met, she will hold a follow-up auction in the next period until some time limit is reached. 
\emph{Potential buyers} decide whether to incur an entry cost to participate in an auction and become a \emph{bidder} as time progresses.
To align with our empirical context, in which the majority of the entry cost comes from the financial constraints and the eligibility screening process, we assume that potential buyers know their private values before they make entry decisions.\footnote{%
A voluminous literature theoretically studies auctions with costly entry. One strand of the literature, which interprets entry costs as information acquisition costs, assumes that potential buyers are uninformed of their values when they make entry decisions (e.g., \citealp{levin1994equilibrium}; \citealp{burguet1996reserve}). Another strand focuses on the costs of preparing bids or achieving eligibility and assumes that potential buyers are fully informed of their values (e.g., \citealp{samuelson1985competitive}; \citealp{mcadams2015benefits}). Recently, a growing literature combines the two modeling approaches and allows potential buyers to receive an arbitrarily informative signal before deciding whether to enter (\citealp{roberts2013should}; \citealp{gentry2017auctions}). This paper follows the second strand and considers fully informed entry, which seems plausible in our empirical application, given that
it is relatively easy for a local resident to inspect the house for sale---while it takes more effort to raise funds to meet the deposit and paid-in-full requirement (see \Cref{subsec:indback} for details of the institutional background).}
We characterize the equilibrium of the recurring auction game with any arbitrary time limit and reserve price sequence. 

Absent costly entry, it is clear from the revelation principle that a single-round auction is optimal for either efficiency or revenue.\footnote{In fact, since potential buyers are ex ante symmetric, any standard auction format (i.e., first-price auction, second-price auction, English auction, Dutch auction) without reserve price achieves fully efficient allocation. With an appropriately chosen reserve price, as in  \citet{myerson1981optimal}, these auction formats maximize revenue.}$^,$\footnote{This implies that price-skimming due to uncertain demand is not why recurring auctions perform better in our setting. We thank a reviewer for suggesting this connection.}
However, with costly entry, recurring auctions give rise to sorted entry and, as a result, dominate single-round auctions. Specifically, in a recurring auction, potential buyers sort their timing of entry according to their private values for the item. Strong potential buyers (who have high values) tend to enter early, while weak potential buyers (who have low values) tend to wait until they have a good chance of winning.
This is because strong potential buyers lose more from waiting, and weak potential buyers' entry costs are more likely wasted if they enter early. 

The sorted entry pattern increases the expected total surplus in two ways.\footnote{\Cref{subsec: ex1} illustrates these two benefits in detail with a simple example.} First, it reduces the probability that two or more bidders simultaneously incur the entry cost, and thereby reduces waste. In particular, weak potential buyers wait until they are sure that others are also not too strong, then enter the auction. This way, their entry costs are less often incurred and wasted in early rounds. 
Second, it increases the probability that the item is ultimately sold. When an auction fails, potential buyers update their beliefs about others' private values. 
In late rounds, they infer that the market is less competitive, since no potential rivals had high enough values to enter in previous rounds. This encourages them to enter the auction, which reduces the likelihood that the item goes unsold. 
Because of these two benefits, recurring auctions with appropriately chosen reserve prices always generate higher expected total surplus than single-round auctions (\Cref{thm: eff}). 
Similarly, we obtain a revenue dominance result (\Cref{thm: rev}). 

We then derive the optimal sequence of reserve prices in a recurring auction.\footnote{This is a non-trivial task, since an explicit solution for the equilibrium is generally unattainable for an arbitrary reserve price sequence. We overcome this challenge by optimizing over entry thresholds directly and then recovering the corresponding reserve prices.}
If the seller aims to maximize efficiency, she trades off between making a marginal potential buyer enter at time $t$ or at time $t+1$. 
Consider a thought experiment in which the marginal potential buyer enters at time $t+1$ instead of time $t$.
Then there is a social gain, since the marginal potential buyer's entry cost is less often incurred because his entry is conditional on no entry by other potential buyers at time $t$. 
However, there are also two kinds of social loss. 
First, there is a loss from discounting because the item may be allocated 1 period later. 
Second, entry at time $t+1$ by other potential buyers could have been avoided had the marginal potential buyer entered at time $t$. 
In the optimum, the choice of reserve prices equalizes the social gain and loss from making a marginal potential buyer enter at time $t+1$ instead of time $t$ (\Cref{thm:eff_design}). 
If the seller aims to maximize profit, she faces a similar trade-off, and the optimal reserve prices can be characterized accordingly (\Cref{thm:rev_design}). In particular, the profit-maximizing condition can be obtained by replacing the marginal potential buyer's value with his virtual value in the efficiency-maximizing condition. 
This is intuitive, because we know from the mechanism design literature that the virtual value accrues to the seller while the value contributes to social surplus. 

Given the theoretical results, it is natural to ask the following quantitative questions in an empirical setting: (i) What are the magnitudes of the efficiency gain and the revenue gain from using a recurring auction relative to a single-round auction? (ii) How much can efficiency and revenue be improved for recurring auctions in practice?

We apply our theory to home foreclosure auctions in China, which represent a significant market. In 2019, over 118,000 foreclosed houses were transacted for 28 billion USD. The market size continues to grow at a high speed, with 149,000 foreclosed houses transacted for 43 billion USD in 2023. 
Using home foreclosure auction data for Fujian province from 2017 to 2019, we estimate structural parameters in a recurring auction model.
The data attest to the important features in our theoretical model setup: Up to three auctions were held in a row for a foreclosed property; the failure rate of the initial auctions was high (38\%), while only 7\% of the foreclosed homes went unsold after all follow-up auctions; and entry costs were considerable due to the financial constraints and the eligibility screening process.\footnote{See \Cref{subsec:indback} for more details.} 

Using simulated maximum likelihood with importance sampling \citep{ackerberg2009new}, our structural estimation comprises two steps. First, for each auction round and each simulation, we solve for the equilibrium entry thresholds using the intertemporal indifference conditions. Second, we calculate the likelihood using the entry thresholds and observable auction outcomes. We contrast the estimation results with those obtained when the intertemporal link between auction rounds is ignored and each auction round is treated as a single-round auction. The recurring auction model performs better in predicting both in-sample and out-of-sample deal prices and numbers of bidders. That the estimation results differ significantly across the two settings affirms the necessity of incorporating the intertemporal link between auction rounds in the empirical analysis. 

Our counterfactual analyses support the theoretical predictions: compared with single-round auctions with the same (first-period) reserve price, recurring auctions raise the annual efficiency by 3.40 billion USD (16.60\%) and revenue by 2.97 billion USD (15.92\%) in China's home foreclosure market.
Using the optimal reserve price sequences derived from our model can further improve efficiency by 0.80 billion USD (3.35\%) and revenue by 0.66 billion USD (3.06\%), respectively. Most of the efficiency and revenue gains are realized by holding a 2-round recurring auction. Increasing the recurring auction rounds from 2 to 3 has a relatively small impact on the auction outcome. 

This paper contributes to the literature on mechanism design with costly entry. \citet{stegeman1996participation} investigates efficiency-maximizing mechanisms in an independent private value (IPV) setting with costly entry. \citet{lu2009auction} and \citet{celik2009optimal} study revenue-maximizing mechanisms within the same framework. 
These studies all focus on single-round mechanisms in the sense that messaging and allocation happen within the same period. 
This single-round assumption is innocuous with free entry. However, it misleads the auction design when entry is costly. Notably, in the single-round mechanism design problem, it is impossible for potential buyers to acquire any information about others' values before deciding whether to enter, because the mechanism only assigns allocation conditional on entry. 
This paper complements prior literature by incorporating the time dimension and considering a dynamic auction setting. 
We demonstrate how sorted entry emerges in recurring auctions, which allows potential buyers to update their beliefs about competitors' values and make more efficient entry decisions. 
\citet{mcafee1997sequentially}; \citet{skreta2015optimal}; and \citet{liu2019auctions} also consider auction games with the possibility of follow-up auctions. However, they abstract away from entry costs and focus on the implications of limited seller commitment on the reserve price sequence and optimal auction design.

Previous literature typically focuses on single-round auctions.\footnote{In a single-round auction in which bids can be submitted or updated over time, dynamic considerations play an important role when rational inattention or bidding frictions---such as limited or stochastic bidding opportunities---exist (\citealp{hopenhayn2016bidding}; \citealp{cho2024explaining}), or when bidders must pay a fee to place or update a bid \citep{dee2020dynamic}. In this case, bidders update their beliefs upon observing previous bids rather than auction failures.} 
The only exception, to our knowledge, is \citet{burguet1996reserve}, in which the authors consider 2-period recurring auctions while assuming uninformed entry and no discounting. Because potential buyers do not know their values the sorted entry pattern does not appear and the benefit whereby weak potential buyers wait to avoid wasting entry costs does not exist. Still, an encouraging entry effect exists in the second round, because potential buyers infer that first-round entrants had bad draws upon observing auction failure. Due to this effect, the authors conclude that a recurring auction can generically achieve greater efficiency and higher revenue than a single-round auction. In the informed entry case, we demonstrate two benefits of the recurring auction, as stated previously: We establish the efficiency and revenue dominance result and provide a complete characterization of optimal reserve price sequences with any number of auction rounds and any discount rate. 

Previous studies have considered other types of dynamics in auction games with costly entry. \citet{ye2007indicative} and \citet{quint2018theory} analyze auction games in which an indicative bidding round precedes the actual auction to screen potential buyers. 
\citet{mcadams2015benefits} studies sequential costly bidding in a second-price auction. 
\citet{bulow2009sellers} and \citet{roberts2013should} also consider sequential costly bidding but fix the order in which potential buyers move. 

We contribute to the empirical literature in two ways. 
First, we introduce and study a new dynamic setting in which multiple auctions can be held across time for the \textit{same} item. 
To our knowledge, this is the first empirical study to incorporate the possibility of follow-up auctions upon auction failure.\footnote{%
Since we consider endogenous entry in recurring auctions, 
this paper is related to a growing empirical literature that considers endogenous entry in single-round auctions (see, e.g., \citealp{li2009entry}; \citealp*{athey2011comparing}; \citealp{krasnokutskaya2011bid}). 
}
Building on previous research on single-round auction estimation (\citealp{laffont1995econometrics}; \citealp{guerre1995nonparametric}; \citealp{guerre2000optimal}; \citealp{hortaccsu2021empirical}; \citealp{luo2024order}), we propose a framework for estimating recurring auctions with costly entry. Our findings suggest that ignoring the dynamics when auctions are recurring can result in underestimating bidders' value distribution and overestimating entry costs. This is because the single-round model overlooks potential buyers' self-selection across time and incorrectly attributes low transaction prices in late auction rounds to low values and high entry costs. 
While a plethora of auction dynamics have been explored in the theoretical literature, related empirical studies are scarce and most focus on repeated sales or the procurement of homogeneous items. 
For instance, in a repeated highway procurement setting, \citeauthor{jofre2000bidding} (\citeyear{jofre2000bidding}, \citeyear{jofre2003estimation}) and \citet{groeger2014study} empirically investigate the effects of backlog or experience on bidders' behavior. \citet{jeziorski2016dynamic} demonstrate that subcontracting allows suppliers to control their capacities more flexibly and thus alleviates dynamic concerns. 

Second, we contribute to the literature by structurally analyzing foreclosure auctions. 
Prior studies focus on price discounts \citep{campbell2011forced, conklin2023alternative}; negative spillover effects on neighborhoods \citep{anenberg2014estimates}; and the causes of foreclosure \citep{corbae2015leverage}, and typically employ a reduced-form approach. 
This paper is one of the first to apply a structural approach to the study of foreclosure auctions. Our estimation enables various counterfactual analyses, and thus informs better designs to promote important social objectives.

The remainder of the paper proceeds as follows. Section 2 lays out the recurring auction model and discusses the single-round benchmark. Section 3 provides some illustrative examples and the theoretical results. Section 4 discusses the data and institutional background of home foreclosure auctions and presents descriptive evidence. Section 5 proposes an empirical framework to estimate the recurring auction model. Section 6 presents the estimation results and Section 7 the counterfactual analyses. Section 8 concludes. 
All proofs are relegated to \Cref{app:proofs}.

\section{Model}\label{sec: model}
A seller has one item for sale, and her value for the item is $v_s$. There are $N$ potential buyers, indexed by the set $\mathcal{N}:=\{1,2,\ldots,N\}$. Each potential buyer $n\in\mathcal N$ has a private value $v_n$ for the item. The values of the potential buyers are independently and identically distributed on the interval $[\underline{v}, \overline{v}]$ according to the cumulative distribution function $F(\cdot)$. 
The seller holds an English auction, or possibly a sequence of English auctions, to sell the item.
We focus on English auctions for ease of exposition and to be consistent with the empirical application, in which the data are from open outcry auctions. 
However, it is worth noting that revenue equivalence can be established in the current setting, so our results hold for all standard auction formats.
Participation or entry in an auction is costly for potential buyers. The entry cost is $K>0$. Potential buyers know their values at the time they make entry decisions. It is natural to require that $v_s<\overline v-K$.

The timing of the auction game is as follows. 
An auction is held at time $t=1$. If no buyer shows up or bids above the reserve price at time $t$,
the seller will hold another auction at time $t+1$ until time $T\geq1$, after which the seller keeps the item forever. 
Before the game starts, the seller sets and publicly announces a reserve price sequence $\bm r:=(r_t)_{1\leq t\leq T}$, where $r_t$ is the reserve price in the auction (possibly) held at time $t$. 
If $T=1$, we have a \emph{single-round auction}; if $T>1$, we refer to the auction as a \emph{recurring auction}.

At time $t$, if an auction is held---which implies that all previous auctions failed---potential buyers simultaneously and independently decide whether to incur an entry cost of $K$ to become a bidder, and bidders who had entered previously are still eligible to bid without incurring the entry cost again. 
For ease of exposition, we assume that potential buyers can only observe the bid history but not the entry history in previous auctions. This implies that if an auction fails, potential buyers cannot tell whether it is due to no entry or no bid above the reserve price.\footnote{As will be seen in \Cref{eqm_analysis}, in the equilibrium we characterize no potential buyer chooses to delay bidding until the next period conditional on entry. With this assumption, we do not need to specify the off-equilibrium-path belief of potential buyers when they see an auction fail due to someone entering without submitting a bid.
If this assumption were dropped and entry history is also publicly observable, we could still obtain the same equilibrium characterization with appropriately specified off-equilibrium-path belief.}
Once the entry decisions are made, the auction begins and a price clock ascends continuously from the reserve price. At the beginning of the auction, each bidder observes the number of all bidders.
At every instant, bidders decide whether to stay in the auction. 
If no bidder participates or all bidders drop out at the reserve price, the auction fails and the game proceeds to the next period (or ends if this is the last period $T$). Otherwise, the price clock continues to ascend so long as two or more bidders remain in the auction, and stops when only one bidder remains. 
The last bidder remaining in the auction wins the item at the final clock price, and the auction game ends. 

The common discount factor for the seller and potential buyers is $\delta\in(0,1)$. As long as the item has not been sold, the seller derives a flow utility $(1-\delta)v_s$ from the item in each period. 
Everything described above, apart from potential buyers' private values, is common knowledge. 



\begin{assumption}[Symmetric Equilibrium]
\label{assum1}
\normalfont
Given that the potential buyers are ex ante symmetric, throughout the paper we restrict our attention to the symmetric Perfect Bayesian Equilibrium (PBE) of the auction game, whereby potential buyers employ identical entry and bidding strategies. 
\end{assumption}

In the equilibrium analysis detailed in \Cref{eqm_analysis}, we show that in equilibrium, bidders will bid truthfully (i.e., a bidder will stay in the auction until the running price reaches his value) and that potential buyers follow a cutoff entry strategy: There is a sequence of entry thresholds $(v_t^*)_{1\leq t\leq T}$, such that at time $t$ a potential buyer will enter the auction if and only if his value is above $v_t^*$. In particular, this means that even though paying the entry cost grants a potential buyer access to the current and all subsequent auctions, the potential buyer bids above the reserve price immediately after incurring the entry cost. 


We take the single-round auction as a benchmark because it performs well in terms of both efficiency and revenue. \citet{celik2009optimal} show that the single-round auction, with an appropriately chosen reserve price, is optimal among all symmetric single-round mechanisms. That is, with an appropriately chosen reserve price, the single-round auction can maximize either the efficiency or the seller's expected profit. Moreover, the authors provide conditions under which the single-round auction is optimal among all single-round mechanisms, symmetric or not.\footnote{Under certain circumstances, an asymmetric equilibrium in a single-round auction can achieve the highest efficiency or revenue (\citealp{stegeman1996participation}; \citealp{lu2009auction}). We discuss this in more detail at the end of \Cref{subsec: ex1}.}

Notably, a disadvantage of the single-round auction---or any other single-round mechanism---is that potential buyers cannot learn anything about their rivals' values before making entry decisions. This foreshadows an efficiency loss in wasteful entry. In contrast, the recurring auction incorporates the time dimension, which allows potential buyers to make dynamic entry decisions. As such, potential buyers have the opportunity to update their beliefs about others' realized values and to economize on their entry decisions. For example, if a potential buyer observes that no one has participated in previous periods, he would infer that others do not have high values for the item and thus feel more confident about entry in the auction. This comparison underlies the superior performance of the recurring auction. We elaborate more on this point in the following section.

\section{Illustrative Examples and Equilibrium Analysis}
\subsection{Illustrative Examples}
\label{subsec: ex1}
Before we analyze the general model in detail, it will be beneficial to illustrate the main insights using simple examples.

\begin{example}\label{ex1}
\textup{
There are $N=2$ potential buyers whose values are independently and identically drawn from the uniform distribution on $[0,1]$. The seller's value is $v_s=0$. A potential buyer's entry cost is $K=0.2$. The common discount factor is $\delta = 0.97$.
}
\end{example}

In this example, there is a unique equilibrium of the single-round auction with no reserve price (\citealp{tan2006equilibria}). 
In the equilibrium, potential buyers use a cutoff entry strategy: Potential buyer $n$ participates and bids $v_n$ in the auction if and only if $v_n\geq 0.45$.\footnote{The cutoff $0.45$ is obtained by solving the indifference equation for the marginal type potential buyer: $v{[F(v)]}^{N-1}-K=0$.}
This equilibrium maximizes the expected total surplus among all single-round mechanisms \citep{stegeman1996participation}. The expected total surplus in this case is 0.39.  

Now consider a 2-period recurring auction (i.e., $T=2$) with a reserve price sequence $(r_1,r_2)=(0.14,0)$.\footnote{This reserve price sequence maximizes the expected total surplus across all reserve price sequences.}
As previously stated, bidders will bid truthfully in equilibrium conditional on entry. To characterize the equilibrium, we only need to pin down entry thresholds. Intuitively, a strong potential player (who has a high realized value) tends to enter early while a weak player tends to wait, for two reasons. First, the strong player loses more if he waits and the item is bought by other potential buyers; he also loses more to discounting if he wins in the next period compared with winning now. And second, the weak player has a lower chance of winning in the current period, which means the entry cost is more likely wasted. We therefore consider the following cutoff entry strategy: Potential buyers whose values are above $v_1^*$ enter in the first period; those with values in between $v_1^*$ and $v_2^*$, with $v_2^*\leq v_1^*$, enter in the second period; and those whose values fall below $v_2^*$ never enter. Equilibrium entry cutoffs can be pinned down by the indifference conditions of marginal potential buyers: A potential buyer with a cutoff value should be indifferent between entering now or in the next period (or, if the current period is the last period, not entering at all). In \Cref{ex1}, we have that $(v_1^*,v_2^*)=(0.66, 0.36)$. The expected total surplus in this case is 0.42, which is higher than that in the single-round auction. 

To understand the reason for the comparison in efficiency, it is useful to note that the recurring auction allows potential buyers to sort their entry over time. This provides an opportunity for players to condition their entry on information updates about potential rivals. 
Sorting generates two benefits for the total surplus: First, weak potential buyers can wait until they are sure that others are also not too strong and then enter the auction. This way, their entry costs are wasted less often. We refer to this as the \emph{economizing on entry} benefit. Second, when an auction fails, potential buyers update their beliefs about the others. They infer that the market is less competitive, since no potential rivals have a high enough value to have entered in the previous period.\footnote{In this example, after the first auction fails, potential buyers know that no one has a value above $v_1^*=0.66$.} Thus they would be encouraged to enter the auction, which reduces the possibility that the item goes unsold. This is reflected by the fact that the entry threshold in the single-round auction (0.45) is higher than the entry threshold in the recurring auction at $t=2$ (0.36). In the single-round auction, the item remains unsold with a probability of 0.2; in the recurring auction, the probability is reduced to 0.13. We refer to this as the \emph{reducing auction failure} benefit. 

\begin{figure}[ht!]
    \centering
    \caption{Equilibrium Entry in Recurring and Single-round Auctions.}\label{ex1_entry}
    \subfloat[Entry in the Single-round Auction.]{
        \includegraphics[width = 0.43\textwidth]{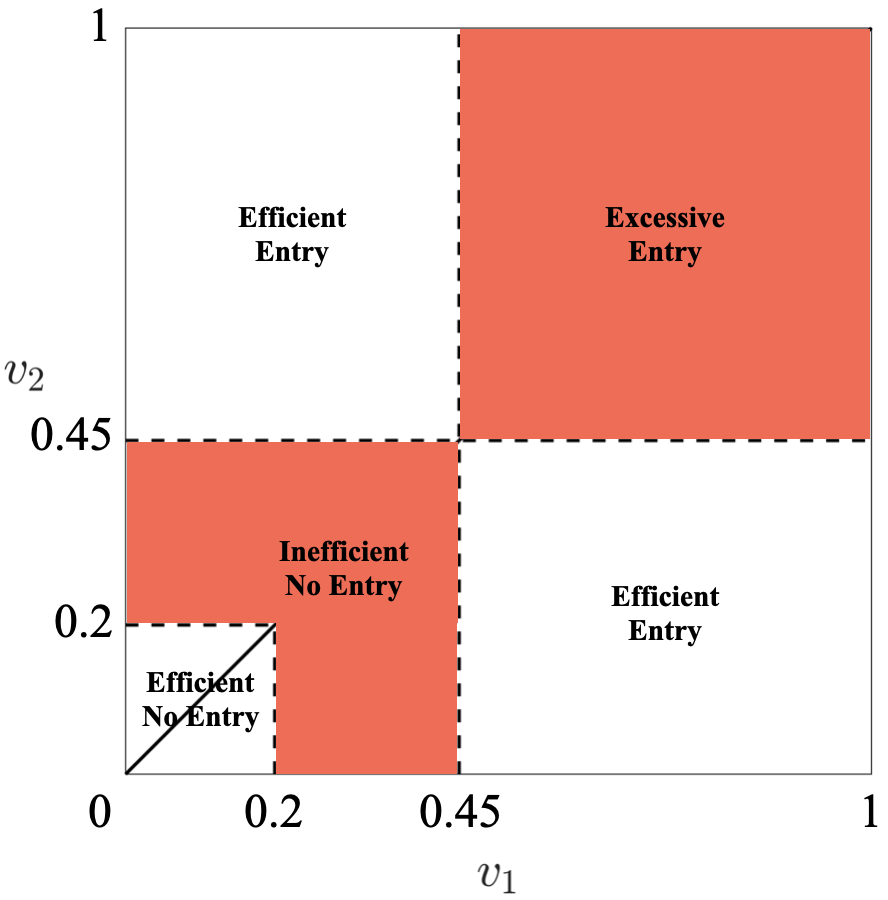}
    }
    \subfloat[Entry in the Recurring Auction.]{
        \includegraphics[width = 0.43\textwidth]{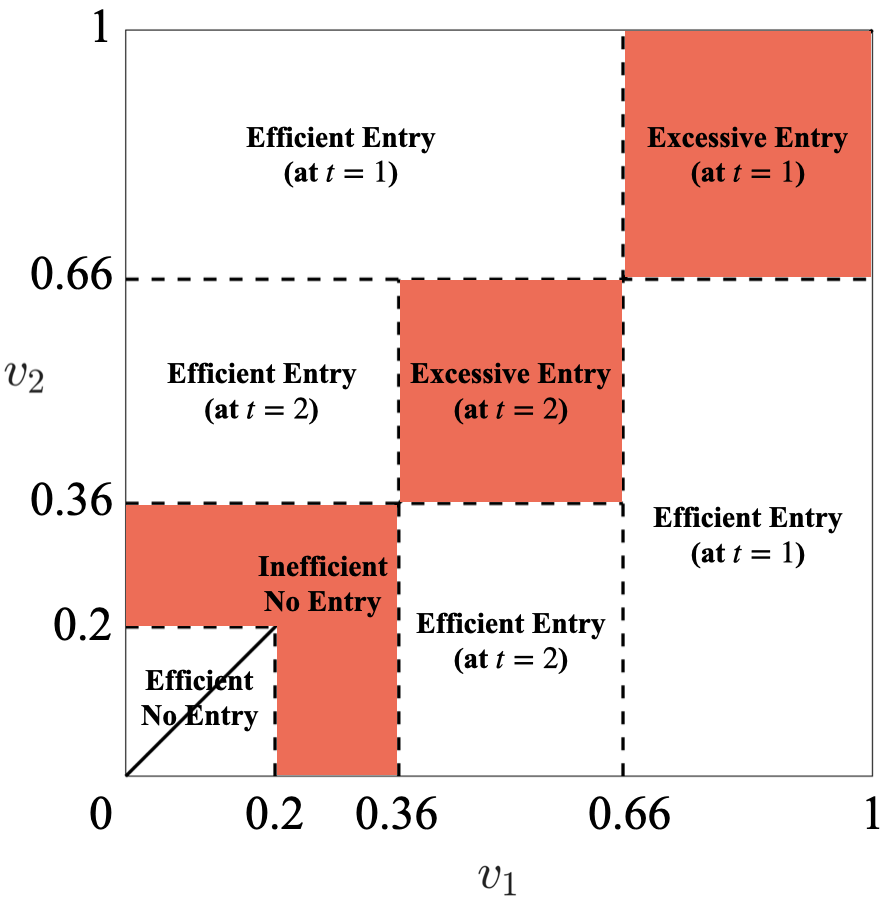}
    }
\end{figure}

We use \Cref{ex1_entry} to visualize the two benefits. \Cref{ex1_entry}(a) shows the entry decisions in the single-round auction. When both potential buyers' values are above the entry cutoff 0.45, there is excessive entry: \emph{Ex post} efficiency would be maximized if the strong potential buyer is the only participant. Value profiles that correspond to excessive entry are denoted by the shaded area in the upper right corner. When both potential buyers' values are below the entry cutoff, but one has a value higher than the entry cost (the shaded area close to the lower left corner), there is no entry, which is not efficient: The total surplus would be raised if the strong potential buyer were to enter. In all other unshaded areas, (no) entry is efficient in equilibrium. 
We similarly plot the efficient and inefficient areas in terms of entry for the recurring auction in \Cref{ex1_entry}(b). Because players sort their entry over time, both the areas that denote excessive entry and inefficient no entry are much smaller than in the single-round auction case.\footnote{When the item is not allocated in the first period, there is a loss in total surplus from discounting---but that does not offset the gain in more efficient entry.} The difference between the excessive entry areas reflects the \emph{economizing on entry} benefit, and the difference between the inefficient no entry areas reflects the \emph{reducing auction failure} benefit.

Notably, as pointed out by \citet{samuelson1985competitive}, setting the reserve price equal to the seller's valuation maximizes the expected total surplus in the single-round auction. In contrast, to maximize efficiency in the recurring auction, a non-trivial reserve price at $t=1$ would be used. This is because the first-period reserve price affects the extent of sorting. If $r_1$ is 0, too many potential buyers would participate in the first period, and this reduces the informativeness of observing whether the first auction has failed. As a result, the gain from sorting is not maximized. In \Cref{ex1_entry}(b),  varying $r_1$ effectively changes the entry threshold in the first period and thus the areas of the two shaded squares. If the entry threshold is too low (resp. too high), the excessive entry at $t=1$ (resp. $t=2$) area will grow too big and squeeze out the \emph{economizing on entry} benefit. 

    

Next, we consider the revenue comparison between the \emph{revenue-maximizing} single-round mechanism and the recurring auction. Because the recurring auction outperforms single-round mechanisms in terms of efficiency, one might expect the recurring auction (with appropriately chosen reserve prices) to also generate more revenue for the seller---after all, the seller's revenue is part of the total surplus.
\citet{celik2009optimal} set up the single-round mechanism design problem with costly entry and study revenue maximization. From their results, it follows that a second-price auction with a reserve price of 0.35 maximizes the revenue across all single-round mechanisms in \Cref{ex1}. The maximized revenue is 0.25. In contrast, with the reserve price sequence set to $(r_1,r_2)=(0.4,0.37)$, the recurring auction generates a revenue of 0.26.

\begin{figure}[ht!]
    \centering
    \caption{Maximized Expected Total Surplus in Single-round and Recurring Auctions While Varying $N$.} 
    \label{fig:eff_N}
    \includegraphics[width=0.5\textwidth]{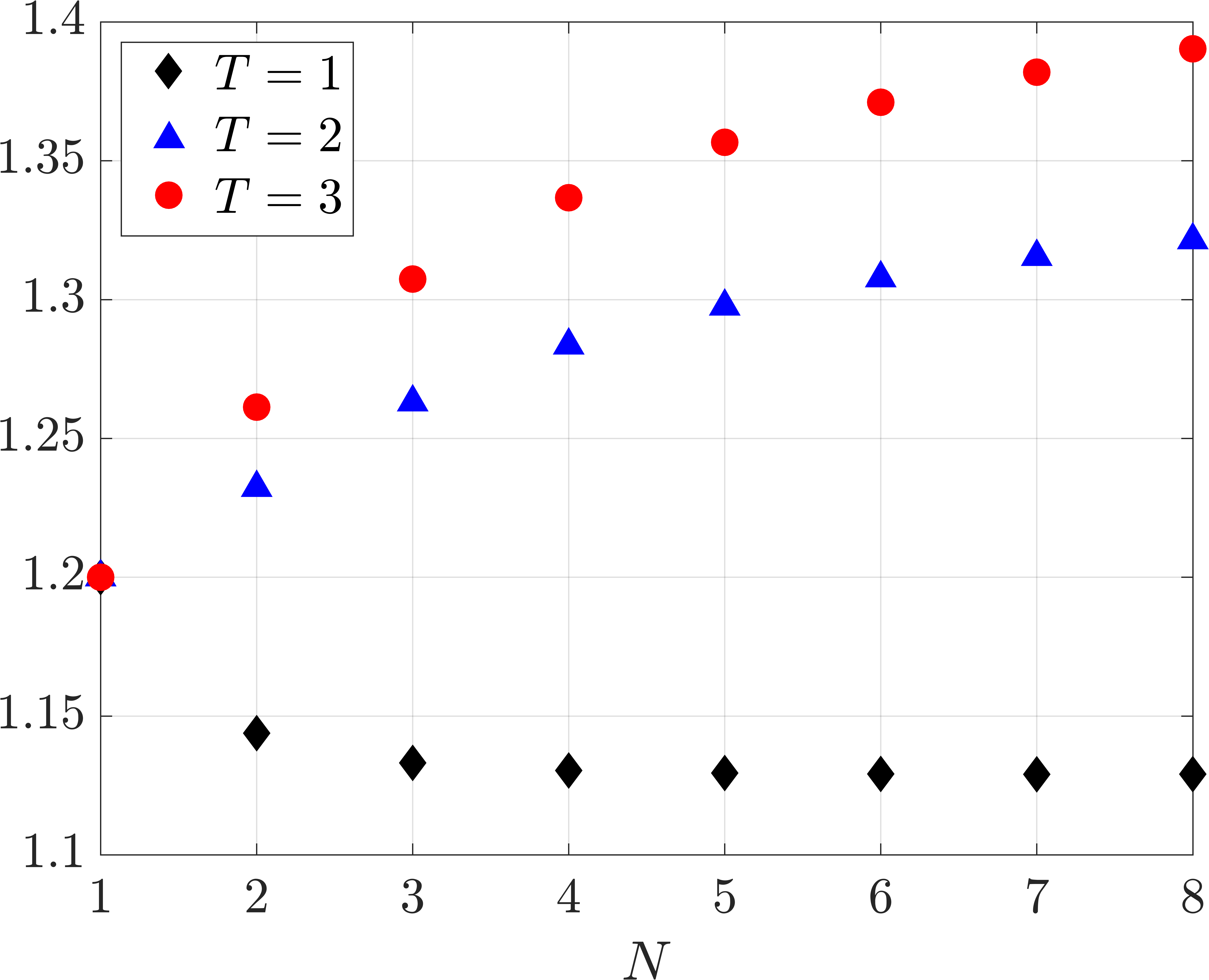}
\end{figure}

Finally, we briefly discuss another example to further illustrate the recurring auction's efficiency dominance over the single-round auction. With costly entry, it is well-known that the efficiency may not be monotonically increasing in the number of potential buyers \citep{samuelson1985competitive}. 
This is because with more potential buyers, each potential buyer worries more about prospective competition and is more inclined to skip the auction to save entry costs.
We demonstrate that recurring auctions can reverse this counterintuitive result using the following example taken from \citet{stegeman1996participation}. 
\begin{example}\label{ex2}
\textup{
There are $N\geq1$ potential buyers whose values are independently and identically drawn from the uniform distribution on $[1,2]$. The seller's value is $v_s=0$. A potential buyer's entry cost is $K=0.3$. The common discount factor is $\delta = 0.97$.
}
\end{example}
The diamond markers in \Cref{fig:eff_N} show how the expected total surplus in the single-round auction changes with the number of potential buyers.\footnote{Recall that a reserve price equal to $v_s=0$ maximizes the expected total surplus in the single-round auction.} It can be seen that the efficiency actually \emph{decreases} with more potential buyers. This reflects the fact that coordinating and economizing on entry is difficult in a single-round setting. Without coordination, the prospect of excessive competition distorts entry and harms efficiency. 
The triangle and circle markers show the maximized expected total surplus (across all reserve price sequences) in 2- and 3-period recurring auctions, respectively. As potential buyers sort their entry over time in recurring auctions, the efficiency is monotonically increasing in the number of potential buyers. 

\citet{stegeman1996participation} points out that coordination may occur in single-round auctions in the form of asymmetric equilibria. Consider the $N=2$ case: Apart from the symmetric equilibrium where a potential buyer enters if his value is above 1.24, there is also an asymmetric equilibrium where one bidder always enters and the other enters if his value is above 1.77. The expected total surplus is 1.22 in the asymmetric equilibrium,\footnote{This is the maximum efficiency that can be achieved by any single-round mechanism.} which is greater than that in the symmetric equilibrium, 1.20. However, relying on asymmetric equilibria to coordinate entry and improve efficiency is precarious, since an asymmetric equilibrium need not always exist.\footnote{In \Cref{ex1}, there is no asymmetric equilibrium in cutoff strategies. More generally, \citet{tan2006equilibria} provide a sufficient condition for there to be a unique symmetric equilibrium and no asymmetric equilibrium in the class of cutoff strategies equilibria.} Moreover, coordination in the form of asymmetric equilibrium may not be very effective compared with the sorted entry pattern in recurring auctions. As \Cref{fig:eff_N} shows, in the $N=2$ case, efficiencies in the 2- and 3-period recurring auctions are 1.23 and 1.26, respectively. 

In a similar vein, \citet{celik2009optimal} and \citet{lu2009auction} provide examples in which asymmetric equilibria in single-round auctions generate higher expected revenue for the seller. However, these improvements are insignificant---and once dynamics are incorporated, the symmetric equilibria in recurring auctions lead to much more sizable improvements.%
\footnote{In Example 1 of \citet{celik2009optimal}, there are two potential buyers whose valuations are distributed according to $F(v)=v^4$ on $[0, 1]$, and the participation cost is $K=0.4$. In the revenue-maximizing single-round auction, the entry cutoffs are asymmetric. A cutoff pair (0.816,0.92) generates a profit of 0.2525 for the seller, whereas the optimal symmetric cutoff, 0.868, generates a profit of 0.25155. However, suppose $\delta=0.97$. The seller's expected revenue is 0.2935 in (the symmetric equilibrium of) a 2-period recurring auction with optimally selected reserve prices.
}$^,$\footnote{
In the example of \citet{lu2009auction}, there are two potential buyers whose valuations are distributed uniformly on $[0.6, 1]$, and the participation cost is $K=0.2$. A cutoff pair (0.66,0.86) generates a profit of 0.431 for the seller, whereas the optimal symmetric cutoff, 0.76, generates a profit of 0.427. However, suppose $\delta=0.97$. The seller's expected revenue is 0.467 in (the symmetric equilibrium of) a 2-period recurring auction with optimally selected reserve prices.
}

\subsection{Equilibrium Analysis}
\label{eqm_analysis}
We now proceed to formal analysis of the model. As stated in \Cref{assum1}, we focus on symmetric PBE of the auction game. In what follows, we show that potential buyers' equilibrium strategy consists of a cutoff entry structure and truthful bidding conditional on entry. 
For notational ease, we denote the CDF of the highest value among $N-1$ potential buyers by $G(\cdot)$. That is, 
$G(\cdot):=[F(\cdot)]^{N-1}.$

Consider a sequence of entry thresholds $\bm v^\dagger:=(v_t^\dagger)_{1\leq t\leq T}$ that satisfy $\overline v\geq v_1^\dagger\geq\ldots\geq v_T^\dagger\geq \underline v$ and focus on a representative potential buyer $n$ with value $v_n$.\footnote{It is convenient to let $v_0^\dagger:=\overline v$.}
Suppose that all the other $N-1$ potential buyers follow the cutoff entry strategy (i.e., for all $1\leq t\leq T$, entering the period-$t$ auction if and only if having a value greater than $v_t^\dagger$) and bid truthfully conditional on entry. Then potential buyer $n$'s interim expected payoff from entering the time-$t$ auction (conditional on this auction being held) and bidding truthfully is $\Pi_t(v_n;\bm v^\dagger)/G(v_{t-1}^\dagger)$, where
\begin{equation}
\Pi_t(v;\bm v^\dagger) := \delta^{t-1} \left[
\int_{v_t^\dagger}^{\min\{\max\{v,v_t^\dagger\},v_{t-1}^\dagger\}}(v-x)dG(x)+(v-r_t)G(v_t^\dagger)-G(v_{t-1}^\dagger)K
\right]. \label{eq_interperiod}
\end{equation}
 
We further specify the sequence of entry thresholds that will render the marginal type $v_t^\dagger$ indifferent between entering at time $t$ or waiting to enter in the future. 
\begin{definition}
\label{def:v}
\normalfont
    Define $\bm v^*:=(v_t^*)_{1\leq t\leq T}$ with $\overline v\geq v_1^*\geq\ldots\geq v_T^*\geq \underline v$ as the solution to 
    $\Pi_t(v_t^*;\bm v^*) \leq \max_{\tau\geq t+1}\Pi_{\tau}(v_t^*;\bm v^*)$ with the equality holding if $v_{t-1}^*>v_{t}^*$.\footnote{We let $\Pi_{T+1}(\cdot)=0$, since there will be no follow-up auctions after time $T$.}
\end{definition}
Intuitively, $v_{t-1}^*=v_{t}^*$ means that the period-$t$ auction is skipped for sure (e.g., due to high reserve price in that period). If no period is skipped, the marginal type $v_{t}^*$ is indifferent between entering at time $t$ or waiting to enter at time $t+1$. In this case, the indifference conditions are simplified to
\begin{equation}\label{eqn:indiff}
    G(v_t^*)(v_t^*-r_t)-KG(v_{t-1}^*)=
    \begin{cases}\delta\left[
    G(v_t^*)(v_t^*-K)-\int_{v_{t+1}^*}^{v_t^*}xdG(x)-r_{t+1}G(v_{t+1}^*)
    \right],&\text{\hspace{-1em} if }t<T,\\
    0,&\text{\hspace{-1em} if }t=T.
    \end{cases}
\end{equation}


We are now ready to present the symmetric PBE.

\begin{proposition}[Equilibrium Characterization]
\label{prop: eqm}
The symmetric PBE of the auction game can be described as follows. At time $t$, if an auction is held, potential buyer $n\in\mathcal N$ chooses to enter if and only if $v_n>v_t^*$. Upon entry, potential buyer $n$ bids truthfully. 
\end{proposition}

With the equilibrium characterization, we now compare the efficiency in recurring auctions and that in single-round auctions. 
Because of the \emph{economizing on entry} benefit and the \emph{reducing auction failure} benefit laid out in \Cref{subsec: ex1}, the following ensues. 

\begin{theorem}[Efficiency Dominance]
\label{thm: eff}
Suppose that $N\geq2$ and $K>0$. Then a recurring auction with an appropriately chosen reserve price sequence will achieve strictly higher efficiency than the symmetric equilibrium in any single-round auction. 
\end{theorem}

Further, we show that the dominance of recurring auctions over single-round auctions also applies to the seller's expected profit. 
\begin{theorem}[Revenue Dominance]
\label{thm: rev}
Suppose that $N\geq2$ and $K>0$. Then a recurring auction with an appropriately chosen reserve price sequence will generate strictly higher profit for the seller than the symmetric equilibrium in any single-round auction. 
\end{theorem}


\subsection{Recurring Auction Design}
\label{subsec:design}
In this part, we study the design of recurring auctions to maximize either efficiency or the seller's profit. It is useful to note that the equilibrium conditions in \Cref{prop: eqm} establish a mapping between reserve price sequences and the entry threshold sequences. The mapping allows us to reformulate the problem of choosing a reserve price sequence to that of choosing an entry threshold sequence. The reformulation will greatly simplify things, since it is generally infeasible to obtain an analytical solution of the entry threshold sequence given an arbitrary reserve price sequence, but the reverse is easy. In fact, an arbitrary entry threshold sequence $\bm v^*$ that is weakly decreasing can be induced by the reserve price sequence $\bm r(\bm v^*)$, with $r_t(\bm v^*)$ given by the following:\footnote{If a certain period is skipped for sure, say $v_{t_0}^*=v_{t_0+1}^*$ for some $1\leq t_0\leq T$, multiple reserve prices can be used for that period. However, they all lead to the same equilibrium outcome.}
\begin{equation}\label{eqn:rofv}
   r_t(\bm v^*) = 
    \frac{1}{G(v_t^*)}\left[
    \begin{array}{c}
         \sum_{\tau=t}^T(1-\delta)\delta^{\tau-t}G(v_\tau^*)v_\tau^*-KG(v_{t-1}^*)+\\
         \sum_{\tau=t}^{T-1}\delta^{\tau-t+1}\int_{v_{\tau+1}^*}^{v_\tau^*}xdG(x)+\delta^{T-t+1}G(v_T^*)v_T^* 
    \end{array}
    \right].    
\end{equation}

We now present and solve the efficiency and revenue maximization problem with respect to entry thresholds.
\paragraph{Efficiency Maximization} 
Given a weakly decreasing entry threshold sequence $\bm v^*$, 
the expected total surplus in the recurring auction is
\begin{equation}\label{eqn:eff}
TS(\bm v^*) = 
\sum_{t=1}^T\delta^{t-1}\bigg\{
\int_{v_t^*}^{v_{t-1}^*}(x-v_s)d[F(x)]^N
-
N[F(v_{t-1}^*)]^{N-1}[F(v_{t-1}^*)-F(v_t^*)]K
\bigg\}+v_s,
\end{equation}
where $\int_{v_t^*}^{v_{t-1}^*}(x-v_s)d[F(x)]^N$ is the expected gain in allocative efficiency in period $t$, and the second term in the bracket is the expected entry cost in period $t$. The efficiency maximization problem is
\begin{equation}\label{eqn:eff_problem}
    \max_{\bm v^*}~TS(\bm v^*)~\text{s.t.}~v_{t-1}^*\geq v_t^*\text{ for all }1\leq t\leq T.
\end{equation}
The key to solving the problem is to show that the constraint $v_{t-1}^*\geq v_t^*$ does not bind. The reason $v_{t-1}^*> v_t^*$ in the optimum can be illustrated by discussing two cases. First, if some final periods are skipped---i.e., $v_{t_0-1}^*>v_{t_0}^*=v_{t_0+1}^*=\ldots=v_T^*$ for some $t_0$---then the seller is effectively holding a single-round auction at time $t_0-1$, which cannot be efficient by \Cref{thm: eff}. Second, if some periods other than the final periods are skipped---i.e., $v_{t_0}^*=v_{t_0+1}^*>v_T^*$ for some $t_0$---then the seller could in effect move every period after $t_0$ to 1 period earlier by using the entry threshold sequence $\bm v^{*'}=(v_1^*,\ldots,v_{t_0}^*,v_{t_0+2}^*,\ldots,v_T^*,v_T^*)$. This is more efficient than the original entry threshold sequence $\bm v^{*}$, since the efficiency gains in periods after $t_0$ are realized earlier. Having established that the constraint does not bind, the solution to \eqref{eqn:eff_problem} is characterized by the first-order conditions. Formally, we have the following result.
\begin{theorem}[Efficient Recurring Auction]
\label{thm:eff_design}
Suppose the seller aims to maximize the expected total surplus by selecting a reserve price sequence of the recurring auction. Then in the efficient design, the equilibrium entry thresholds $\bm v^*$ solve the following:
\begin{equation}\label{eqn:eff_foc}
\begin{cases}
    \left\{\left[\frac{F(v_{t-1}^*)}{F(v_t^*)}\right]^{N-1}-\delta\frac{NF(v_t^*)-(N-1)F(v_{t+1}^*)}{F(v_t^*)}\right\}K-(1-\delta)(v_t^*-v_s)=0,&\text{ if }t<T,\\
    \left[\frac{F(v_{t-1}^*)}{F(v_t^*)}\right]^{N-1}K-(v_t^*-v_s)=0,&\text{ if }t=T.
\end{cases}
\end{equation}
The corresponding reserve price sequence can be obtained from \eqref{eqn:rofv}.
\end{theorem}

To see the intuition behind \Cref{thm:eff_design}, let us consider the social gain and loss of making the marginal potential buyer with value $v_t^*$ enter at time $t+1$ instead of time $t$. 
The social gain manifests as saving in entry costs. Conditional on time $t$ being reached (i.e., no one has shown up in previous auction rounds), the cost of participating at time $t$ is $K$. However, if the marginal buyer participates at time $t+1$, he incurs entry cost only if no one shows up at time $t$, so the expected entry cost is $\delta[F(v_t^*)/F(v_{t-1}^*)]^{N-1}K$. Consequently, the social gain of cost saving is $\{1-\delta[F(v_t^*)/F(v_{t-1}^*)]^{N-1}\}K$. 
The social loss consists of two parts. First, in the event that all other potential buyers have values below $v_t^*$, there is a loss of allocating the item 1 period later. Specifically, the loss from discounting is $(1-\delta)[F(v_t^*)/F(v_{t-1}^*)]^{N-1}(v_t^*-v_s)$. Second, if the marginal potential buyer had entered in period $t$, entry by other potential buyers in period $t+1$ could have been avoided. 
Since there are $N-1$ other potential buyers and the probability of each of them entering in period $t+1$, conditional on period $t+1$ being reached, is $[F(v_t^*)-F(v_{t+1}^*)]/F(v_t^*)$, 
the waste of period-$(t+1)$ entry is $\delta[F(v_t^*)/F(v_{t-1}^*)]^{N-1}(N-1) [F(v_t^*)-F(v_{t+1}^*)]/F(v_t^*)K$. 

To maximize the expected social surplus, the choice of entry thresholds must equalize the social gain and social loss. 
We rewrite the condition in \Cref{thm:eff_design} for the $t<T$ case as follows to reflect the trade-off.
\[
\underbrace{\left\{1-\delta\left[\frac{F(v_t^*)}{F(v_{t-1}^*)}\right]^{N-1}\right\}K}_{\text{social gain in entry cost saving}}
=
\underbrace{(1-\delta)(v_t^*-v_s)}_{\substack{\text{social loss from} \\ \text{delayed allocation}}}
+
\underbrace{\delta\left[\frac{F(v_t^*)}{F(v_{t-1}^*)}\right]^{N-1}(N-1)\frac{F(v_t^*)-F(v_{t+1}^*)}{F(v_t^*)}K}_{\text{social loss from period-$(t+1)$ entry}}.
\]

\paragraph{Profit Maximization} 
The problem of maximizing profits requires additional effort, because it is non-trivial to derive the expression of the seller's expected profit in terms of entry thresholds $\bm v^*$. To do that, we first express the seller's expected profit as a function of both entry thresholds and reserve prices: 
\[
    R(\bm v^*;\bm r)=\sum_{t=1}^T\delta^{t-1}\left\{
    \begin{array}{c}
N(N-1)\int_{v_t^*}^{v_{t-1}^*}(x-v_s)f(x)[F(v_{t-1}^*)-F(x)][F(x)]^{N-2}dx\\
+
N[F(v_{t-1}^*)-F(v_t^*)][F(v_t^*)]^{N-1}(r_t-v_s)
\end{array}
\right\},
\]
where the first line in the bracket is the expected profit in the event that at least two bidders show up at time $t$, and the second line in the bracket is the expected profit when exactly one bidder presents at time $t$ (so the reserve price is charged).

Next, we plug $\bm r(\bm v^*)$ given by \eqref{eqn:rofv} into $R(\bm v^*;\bm r)$. After algebraic manipulation, we have that
\begin{equation}
\label{eqn:rev}
\begin{split}
R(\bm v^*;&\bm r(\bm v^*))= \delta^TN[1-F(v_T^*)]G(v_T^*)v_T^*+
\\
&\sum_{t=1}^T\delta^{t-1}
\left\{
\begin{array}{cc}
     N\int_{v_t^*}^{v_{t-1}^*}x[1-F(x)]dG(x)+N(1-\delta)\left[1-F(v_t^*)\right]G(v_t^*)v_t^* \\
     -NG(v_{t-1}^*)[F(v_{t-1}^*)-F(v_t^*)]K
-v_s\left[(F(v_{t-1}^*))^N-(F(v_t^*))^N\right]
\end{array}
\right\}
.
\end{split}
\end{equation}
The profit maximization problem becomes
\begin{equation}\label{eqn:rev_problem}
    \max_{\bm v^*}~R(\bm v^*;\bm r(\bm v^*))~\text{s.t.}~v_{t-1}^*\geq v_t^*\text{ for all }1\leq t\leq T.
\end{equation}
A technique similar to that in the efficiency part would establish that the constraint $v_{t-1}^*\geq v_t^*$ does not bind in the optimum. As a result, \Cref{thm:rev_design} characterizes the profit-maximizing recurring auction. 

\begin{theorem}[Profit-maximizing Recurring Auction]
\label{thm:rev_design}
Suppose the seller aims to maximize her profit by selecting a reserve price sequence of the recurring auction. Then in the profit-maximizing design, the equilibrium entry thresholds $\bm v^*$ solve the following:
\begin{equation}\label{eqn:rev_foc}
\begin{cases}
    \left\{\left[\frac{F(v_{t-1}^*)}{F(v_t^*)}\right]^{N-1}-\delta\frac{NF(v_t^*)-(N-1)F(v_{t+1}^*)}{F(v_t^*)}\right\}K-(1-\delta)\left[v_t^*-\frac{1-F(v_t^*)}{f(v_t^*)}-v_s\right]=0,&\text{ if }t<T,\\
    \left[\frac{F(v_{t-1}^*)}{F(v_t^*)}\right]^{N-1}K-\left[v_t^*-\frac{1-F(v_t^*)}{f(v_t^*)}-v_s\right]=0,&\text{ if }t=T.
\end{cases}
\end{equation}
The corresponding reserve price sequence can be obtained from \eqref{eqn:rofv}.
\end{theorem}

Notably, the condition for profit-maximizing \eqref{eqn:rev_foc} is similar to the condition for efficiency-maximizing \eqref{eqn:eff_foc}. In fact, the profit-maximizing condition can be obtained by replacing $v_t^*$ with the virtual value $v_t^*-[1-F(v_t^*)]/f(v_t^*)$ in the relevant places in the efficiency-maximizing condition. This is intuitive, since the mechanism design literature suggests that the virtual value accrues to the seller while the value contributes to social surplus. 

In what follows, we take our model to a practical setting and empirically estimate the model parameters. The structural estimation allows us to quantitatively identify the efficiency gain and revenue gain brought about by recurring auctions. We also conduct counterfactual analyses to inform better auction design. 

\section{Industry Background, Data, and Descriptive Evidence}
This section introduces the background of the home foreclosure auction market in China and presents the data, summary statistics, and some descriptive evidence.

\subsection{Industry Background}
\label{subsec:indback}
In China, the market for home foreclosure auctions is substantial, with a transaction volume of 196 billion CNY (28 billion USD) in 2019.  Since 2014, it has become common practice for Chinese local courts to conduct foreclosure auctions through online platforms, with over 90\% of these auctions taking place on Alibaba. Since 2017, the supreme court has mandated that auctions for foreclosed properties must be held publicly and online. 

\begin{figure}[ht!]
     \centering
     \caption{Distribution of the Time Gap between Two Consecutive Auctions.}
     \label{fig:time_gap}
 \includegraphics[width = 0.5\textwidth]{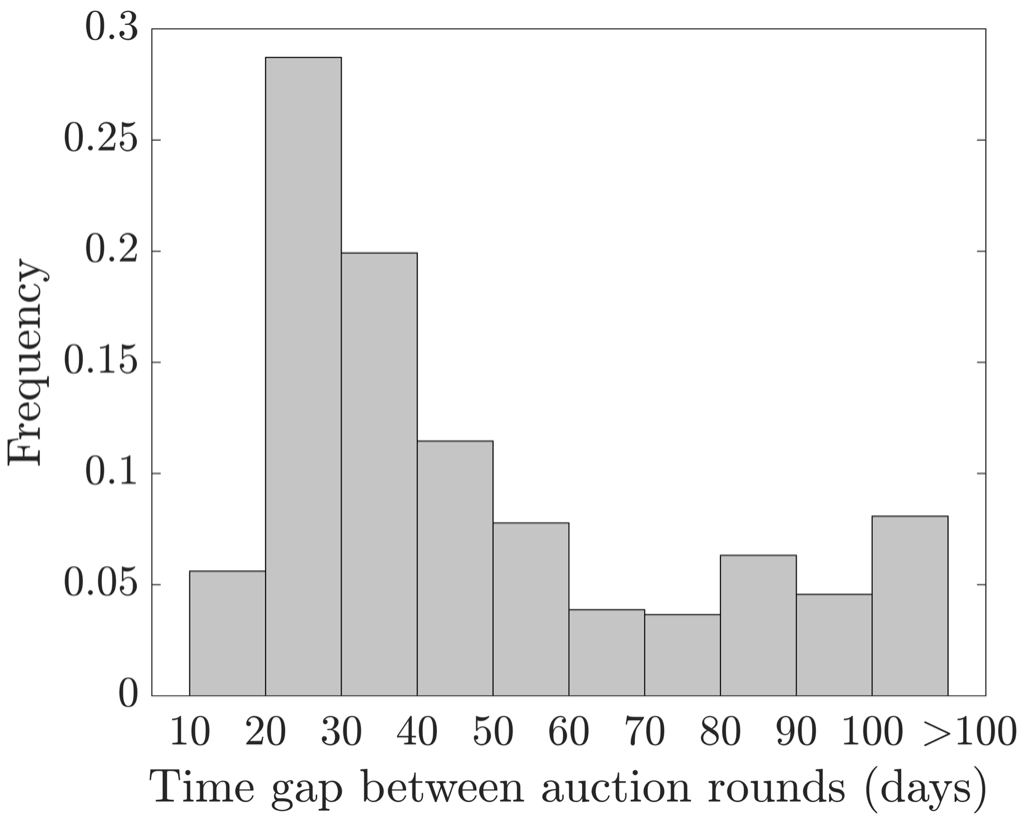}
 \end{figure}

When a foreclosed property is up for auction, the government will commission a company to assess the property's market value. 
This assessed price will be used as a benchmark to determine the reserve price in the auction and any follow-up auctions. In the initial auction, it is required by law that the reserve price is set above 70\% of the assessed price. For more than 60\% of the properties in our sample, the initial reserve price is between 70\% and 80\% of the assessed price. Information regarding the property, the assessed price, and the reserve price will be posted online 1 month prior to the auction date.  An open outcry auction is held on the auction date if any potential buyer qualifies as a bidder and participates. The open outcry stage lasts for 24 hours. During this stage, bidders can observe the standing bid and decide whether to raise their own bid. If no bid (above the reserve price) is received, the auction fails. If a property is not successfully sold in the initial auction, the government will typically hold a follow-up auction within 2 months. The median time gap between two auction rounds in our sample is 36 days. \Cref{fig:time_gap} presents the distribution of the time gap between two auction rounds.  

The law mandates that the reserve price in the second auction is set above 80\% of the reserve price in the initial auction. In practice, the vast majority of second-round reserve prices are set at exactly 80\%.\footnote{In our sample, the average reserve price in the second round is 82\% of the reserve price in the initial auction.} If the second auction also fails, the property will be listed again for the third and \emph{final} auction.\footnote{The third auction is referred to as the liquidation stage by the court. In 2017, there was no difference between this stage and the first two rounds, so the liquidation stage can be treated simply as the third auction round. In 2018, the Supreme Court of Fujian province (where our data are from) changed the auction rule for the liquidation stage. Following the change, a foreclosed property can be listed for sale at the liquidation stage for a 60-day period. If a bid is received during this period, an open outcry auction is initiated, and other qualified bidders can participate within a 24-hour timeframe. Since it is impossible to finish all the paperwork and qualify for participation within 24 hours, bidders cannot condition their entry decisions on others' actions.  For the sample period after the policy change, we continue to treat the liquidation stage as the third round of the recurring auction game. This simplification offers great tractability without much loss. As we shall see, the outcomes of 2-period recurring auctions are similar to those of 3-period auctions, which indicates that the value of having additional rounds beyond the second is limited. As a robustness check, we use the 2017 subsample to estimate the recurring auction model. The results (reported in Online \Cref{estimation_year}) are similar to those obtained from the full sample.} 
According to the law, the reserve price in the third round must remain the same as it was in the second auction.

Entry costs in the auction market are non-negligible. Potential buyers must fulfill several requirements to qualify as bidders, including depositing 10\% to 20\% of the reserve price before the auction to demonstrate their eligibility. It is worth noting that the deposit itself is not an entry cost, since it is refundable. However, interest proceeds (or costs) during the period the deposit is frozen are considered entry costs. In the event a potential buyer has to borrow to meet the deposit requirement, efforts to obtain those funds are also entry costs. Furthermore, the winning bidder must pay the full amount within 5 working days, which poses a financial challenge for home buyers because obtaining a mortgage from a bank within 5 working days is infeasible. Along with the liquidity cost, there are also bureaucratic costs associated with preparing documents to participate in the auction.\footnote{The cost of services that assist with filing the paperwork is approximately 100 USD.}

Purchase and reselling restrictions in the housing market discourage speculators from participating in home foreclosure auctions. Households in most cities in our sample are allowed to own at most one or two houses. 
Reselling the property within 2 years of purchase will result in a value-added tax and surtax totaling 5.6\% of the house's value.
Although firms are not subject to the purchase restriction, they are heavily taxed for owning houses. The annual property tax amounts to 1.2\% of the house value or 60\% of the rent. Also, when a firm resells a house, the combined value-added tax and sales tax exceed 30\% of the value added. 
As a result, participants in home foreclosure auctions are likely to be ordinary buyers rather than speculators.\footnote{In an interview, the owner of Dongfang Hanhai Auction Co., an intermediary that helps clients prepare documents for foreclosure homes, stated that most clients are buying homes for their own residential use.}

The homebuying restrictions and resale taxes mitigate the concern of interdependent valuations resulting from a resale market \citep{haile2001auctions} and support our IPV assumption. Moreover, in our empirical model setup, we allow the mean and variance of the value distribution to depend on observable characteristics of the property, and thereby encompass a common value component. The private value component stems from the fact that people derive different utilities from owning the same house with a particular combination of characteristics.\footnote{The IPV assumption has been widely adopted in various empirical settings. See, e.g., \citet*{athey2011comparing} and \citet{roberts2013should} for timber auctions and \citet{allen2023resolving} for failed-bank assets auctions.}
While we do not formally test the IPV framework against a common value 
alternative,\footnote{%
The existing literature (e.g., \citealp{athey2002identification}; \citealp{bougt2023identification}) typically assumes away costly entry when developing tests of private versus common value models, and thus the tests do not directly apply to our setting. Moreover, conducting the tests requires observing more than the winner's bid. However, the bid data we observe from open outcry auctions cannot simply be treated as bid data from standard auctions (see \cref{fn:wb}). This data limitation further hinders our ability to perform a formal test.} 
one piece of suggestive evidence in favor of the former is that in our sample, bidders rarely wait until near the end of the auction to submit their bids,\footnote{Only around 8\% of the auctions receive initial bids within the last 10 minutes. In most auctions, initial bids are submitted at the very beginning.} which would have been optimal if common value is a concern (\citealp{bajari2003winner}) because bidders would not want opponents to learn about their values. 

\subsection{Data}
\label{subsec:data}
The home foreclosure auction data used in this paper come from CnOpenData, a data consulting company. CnOpenData has collected information on home foreclosure auctions via Alibaba's digital platform since 2017 \citep{CnOpenData}. The dataset contains detailed information about each auction, including a property identification number, the deal price (if any), the reserve price, the number of bidders, and the bidding records of each bidder.\footnote{\label{fn:wb}Although we observe the bidding records of all bidders, it is known that the losing bids in open outcry auctions may be lower than the bidders' values \citep{haile2003inference}. Therefore, in our estimation, we use only the winning bid and take it as the second-highest realized value.} We also observe the assessed value, area, and location of the property being sold.

Our data consist of home foreclosure auctions that took place in Fujian province from 2017 to 2019. The province's population is approximately 41 million, roughly equivalent to that of California. Its per capita GDP is 15,500 USD. Our sample includes a total of 11,675 \emph{auctions} for 7,973 \emph{foreclosed homes}.\footnote{In our analysis, we focus on foreclosed homes and exclude other types of foreclosed properties, such as parking spaces, shops, commercial apartments, and factories, from our sample.}
During the sample period, the total transaction volume of home foreclosure auctions in Fujian is 9.2 billion CNY, or 1.3 billion USD. 

Table \ref{tab_sumrounds} presents the recurring pattern of home foreclosure auctions. The auction failure rate stands at 38\% in the first round and becomes 31\% in the second round. It increases to 68\% in the third and final round, possibly because there is no reserve price reduction from the second to the third round. As can be seen in the table, there is some attrition across auction rounds. In the first round, 3,002 houses were unsold, of which 150 houses never appeared in subsequent auctions. In the second round, 889 auctions failed, and 850 of those houses entered the third stage afterward. Overall, the attrition rate is around 5\%. Attrition between auction rounds may be caused by various reasons, such as the foreclosed property owner securing enough funds to pay off their outstanding debts, legal complications arising for either the lender or the foreclosed property owner, or the lender and owner reaching an agreement.\footnote{In Online \Cref{subsec:balance}, we show that the observable characteristics of the houses that experience attrition do not differ significantly from those of the other houses.}
We incorporate the possibility of houses ``disappearing'' when setting a discount factor. For our empirical analysis, we exclude unsold properties that disappeared from the market before the third round.\footnote{11,411 auctions for 7,772 foreclosed homes remain after removing attrited properties.}

\begin{table}[ht!]
\begin{center}
\begin{threeparttable}
  \centering
  \caption{Number of Auctions across Rounds.} \label{tab_sumrounds}
    \begin{tabular}{lccc}
    \toprule
          & \multicolumn{1}{l}{First auction} & \multicolumn{1}{l}{Second auction} & \multicolumn{1}{l}{Third auction} \\
    \midrule
    Success & 4971   & 1963   & 271 \\
    Fail  & 3002   & 889    & 579 \\
    Total & 7973   & 2852   & 850 \\
    Auction failure rate & 0.38 & 0.31 & 0.68\\
    \bottomrule
    \end{tabular}%
    \end{threeparttable}
    \end{center}
\end{table}%



Echoing our theoretical analysis and for tractability, we assume that the same set of potential buyers are interested in a foreclosed house across time. 
Since the number of potential buyers cannot be directly observed, we construct a proxy using the number of individuals who bookmarked and browsed the properties before they are auctioned. Specifically, the number of potential buyers is defined as the number of individuals who have shown interest in the first auction round, divided by 1,000, plus the number of entrants.\footnote{Our results are robust to alternative definitions of potential entrants using factors other than 1/1,000. Estimation and counterfactual analysis results for factors 1/500 and 1/1,500 are reported in Online \Cref{subsec: N robustness}.}$^, $\footnote{Notably, differences in the number of bookmarks between auction rounds are small. For houses that failed to sell in the initial auction, the average number of bookmarks in the following round is only about 4\% higher.}

To further alleviate concerns about the set of potential buyers or their valuations changing over time, we estimate the empirical model for a subsample with short time lags (bottom quartile, or less than 24 days) between the first two auction rounds and a subsample with long time lags (top quartile, or more than 44 days). The estimation results are reported in Online \Cref{estimation_lag}.
Also, we estimate a single-round auction model, which corresponds to a new set of potential buyers arriving in each period while previous potential buyers exit. This alternative specification performs worse than the recurring auction specification (with a constant set of potential buyers) in model fit. 

Table \ref{tab_summary} presents summary statistics of the main variables used in this paper.
In \Cref{tab_summary}, auctions are pooled together across rounds. The average assessed value for a house is about 1.43 million CNY (210 thousand USD), which is approximately 108 thousand CNY (16 thousand USD) higher than the average deal price.
The average reserve price is 1.12 million CNY (165 thousand USD).
The overall auction success (resp. failure) rate is 63\% (resp. 37\%).

\begin{table}[ht!]
  \begin{center}
      \begin{threeparttable}          
  \caption{Summary Statistics.}   \label{tab_summary}%
     \begin{tabular}{lcccccc}
    \toprule
        & & & \multicolumn{3}{c}{Percentiles} & \\
        \cmidrule(lr){4-6}
          & Mean & Std. dev. & 0.25 & 0.5 & 0.75 & Obs \\
    \midrule
    Deal price (10K CNY) & 132.71 & 123.71 & 56.00 & 88.35    & 157.24 & 7201 \\
    Assessed price (10K CNY) & 143.52 & 144.00 & 59.09  & 90.52 & 162.69 & 11411 \\
    Reserve price (10K CNY) & 111.79 & 119.70 & 44.00 & 69.23    & 124.50 & 11411 \\
    Success & 0.631 & 0.483 &       &       &       & 11411 \\
    Number of bidders & 3.29  & 3.98 & 0     & 1     & 5     & 11411 \\
    Number of potential entrants & 9.58  & 6.01   & 5     & 8     & 13    & 11411 \\
    Area ($m^2$)  & 130.11 & 45.75  & 97.19 & 127.90 & 155.67 & 11411 \\
    Distance to city center (km) & 7.24  & 14.00 & 1.26  & 2.40  & 5.55  & 11411 \\
    \bottomrule
    \end{tabular}%
          \end{threeparttable}
  \end{center}
\end{table}%

The home foreclosure auction market is an exemplary empirical setting for our recurring auction model. In particular, the auction failure rates are high, and the entry costs are sizable. 
Before diving into the structural analysis, we first present some reduced-form evidence of dynamics and intertemporal trade-offs in the market. 

\subsection{Hedonic Housing Price Regression}
We use the following hedonic regression to investigate how various factors affect the deal price in home foreclosure auctions, with special emphasis on the effects of auction rounds:
\begin{equation}\label{eq_sorting}
\begin{split}
    \log(p_i)=\alpha_0&+\alpha_1 \log(assess_i)+ \alpha_2 area_i + \alpha_3 \log(dist_i)\\
    &+ \alpha_4 D_{round=2, i}+ \alpha_5 D_{round=3, i} 
   \alpha_6(\text{\# of potential entrants})_i
    +\eta_t +\epsilon_i.
\end{split}
\end{equation}
Specifically, $p_i$ is the deal price, $assess_i$ represents the assessed price, $area_i$ is the construction area of the property, and $dist_i$ denotes property $i$'s distance to the city center. $D_{round=2, i}$ and $D_{round=3, i}$ are dummies that switch on if the auction round is 2 and 3, respectively. Auctions in the first round are set as the base group. $\text{\# of potential entrants}_i$ is the number of the potential entrants. $\eta_t$ controls for year-by-month fixed effects. Table \ref{tab_sorting} reports regression results with gradually added controls. Column (5) corresponds to the specification in \Cref{eq_sorting}. 

\begin{table}[ht!]
  \begin{center}
      \begin{threeparttable}
        
  \caption{Effects of Auction Rounds on Deal Price.} \label{tab_sorting}
     \begin{tabular}{lccccc}
      \toprule
           & \multicolumn{1}{c}{(1)} & \multicolumn{1}{c}{(2)} & \multicolumn{1}{c}{(3)} & \multicolumn{1}{c}{(4)} & \multicolumn{1}{c}{(5)}\\
           & \multicolumn{5}{c}{log(deal price)} \\
    \midrule
    log(assessed price) & 0.981 & 0.981 & 0.988 & 0.944 & 0.945 \\
          & (0.003) & (0.003) & (0.003) & (0.003) & (0.015) \\
    round=2 &       & -0.204 & -0.199 & -0.135 & -0.135 \\
          &       & (0.005) & (0.005) & (0.004) & (0.004) \\
    round=3 &       & -0.306 & -0.299 & -0.172 & -0.171 \\
          &       & (0.011) & (0.011) & (0.010) & (0.009) \\
    area (100 m$^2$)  &       &       & -0.016 & 0.0284 & 0.0250 \\
          &       &       & (0.005) & (0.004) & (0.005) \\
    log(distance to city center) &       &       & -0.022 & -0.019 & -0.020 \\
          &       &       & (0.002) & (0.001) & (0.001) \\
    \# of potential entrants & & & & 0.019 & 0.018\\
    & & & &  (0.000) & (0.000)\\
    Year-by-month fixed effects &       &       &       &      & X \\
    Observations & 7,201  & 7,201  & 7,201  & 7,201  & 7,201 \\
    $R^2$ & 0.933 & 0.949 & 0.950 & 0.966 & 0.967 \\
    \bottomrule
    \end{tabular}%
\begin{tablenotes} \small
\item[\hspace{-1em}]     Notes: (1) First-round auctions (round=1) are set as the base group, and the corresponding coefficient is absorbed. (2) Standard errors in parentheses. (3) All coefficients are significant at the $p<0.01$ level.
\end{tablenotes}
        \end{threeparttable}
  \end{center}
\end{table}%

The regression results suggest that the assessed price accurately reflects the housing value, since its coefficients are close to 1 in all specifications. Column (1) shows that assessed price alone can explain 93\% of the variation in the deal price. 
Under the full specification, shown in Column (5), the transaction price for a property sold in the second round is 14.4\% (exp(0.135)-1) lower than a property sold in the first round. The price reduction is 18.6\% (exp(0.171)-1) for a property sold in the third round. 
Note that we do not control for the reserve price in the regressions: It is a bad control \citep{angrist2009mostly}, because it is almost perfectly determined by the auction round and the assessed price. 
This means that we cannot econometrically disentangle the effects of reserve price changes and auction rounds on deal prices. 
However, since the reserve price remains the same from the second to the third round, the 4.2\% price drop between these two rounds reflects the auction round effect.

These findings support the sorted entry pattern predicted by our theoretical model. That is, potential buyers with high private values tend to participate in early auction rounds, while weak potential buyers tend to wait and enter late. 
The sorted entry pattern offers an explanation for lower prices in later auction rounds. 
Alternatively, if sorting is absent and each auction independently draws a new set of potential buyers, the auction round should not have a significant impact on the deal price.\footnote{While other factors such as the ``stigma'' effect may also negatively impact auction prices, they are likely to be much smaller in magnitude than the estimates in Table \ref{tab_sorting}. For instance, \cite{cortes2022stench} suggest that the size of the stigma effect in the Australian home auction market is approximately 1\%.}


\section{Empirical Strategy}
In this section, we specify our empirical model, introduce the simulated maximum likelihood method, and briefly discuss the sources of identification.

\subsection{Empirical Model}
Our empirical model is specified as follows. For property $i$, we assume each potential buyer's value follows a truncated log-normal distribution, i.e., $v_n \sim TRLN(\mu_i, \sigma_{i}, \underline v,\overline v)$, where $\mu_i$ and $\sigma_{i}$ are the location parameters of the log-normal distribution,
and $\underline v$ and $\overline v$ are the lower and upper truncation bounds.\footnote{In practice, we set the lower bound $\underline v$ to be sufficiently low and the upper bound $\overline v$ to be sufficiently high, thereby rendering the truncation of the distribution negligible.}
Then the auction setting for property $i$ is described by the vector of parameters $\widetilde\Lambda_i=(\mu_i,\sigma_{i},K_i; N_i,T,\delta,\bm r_i)$, where $K_i$ is the entry cost, $N_i$ is the number of potential buyers, $T$ is the maximum number of auction rounds, $\delta$ is the discount factor, and $\bm r_i=(r_{ti})_{1\leq t\leq T}$ is the reserve price sequence. 

The parameters $(N_i,T,\delta,\bm r_i)$ can be obtained using information from the data. 
As discussed in \Cref{subsec:data}, the number of potential buyers is proxied using the number of individuals who bookmarked and browsed the property online. We set $T=3$ to reflect the institutional setup and let $\delta=0.95$, which encompasses both discounting and the probability of a property's being withdrawn from the market. 
For auctions that actually happened, we can observe the reserve prices. But if a property is sold in either the first period or the second period, we cannot observe the full reserve price sequence, as the subsequent auctions never took place. In such cases, we follow the common practice we see in the data and assume that the second-period reserve price is 80\% of the initial reserve price, and the third-period reserve price is the same as the second-period reserve price. 

Our objective is to identify the value distribution parameters $(\mu_i, \sigma_{i})$ and the entry cost $K_i$ in home foreclosure auctions. For notational ease, we define
$\Lambda_i := (\mu_i, \sigma_{i},K_i).$
Given a set of observables $X_i$, including a constant, the logarithm of the assessed price, the area, and the logarithm of the distance to city center, we assume the auction parameters $\Lambda_i = (\mu_i, \sigma_{i},K_i)$ are governed by the following truncated normal distributions:\footnote{We denote by $TRN(\mu,\sigma,\underline D,\overline D)$ a truncated normal distribution with location parameters $\mu$ and $\sigma$ and truncation bounds $\underline D$ and $\overline D$.}$^,$\footnote{Note that $\mu_i$, $\sigma_{i}$, and $K_i$ are in 10K CNY (1.5K USD). The truncation bounds are chosen to cover sensible ranges of these parameters. For example, with $\mu_i=7$, the mean of the value distribution $TRLN(\mu_i, \sigma_{i}, \underline v,\overline v)$ is above $\exp(7)=1,096$, which is greater than the highest deal price in our sample.
}
\begin{align}
 \mu_i \sim TRN (X_i\beta_\mu, \omega_\mu, 1,7),~
\sigma_i  \sim TRN (X_i\beta_\sigma, \omega_\sigma, 0.01 ,3),~ 
K_i \sim TRN (X_i\beta_K, \omega_K, 0,15), \label{eq_est}  
\end{align}
where $(\beta_\mu, \beta_\sigma, \beta_K)$ are the coefficients of the covariates and $(\omega_\mu,\omega_\sigma,\omega_K)$ are the standard deviations of the untruncated distributions. We denote the set of parameters to be estimated by $B$. Formally, 
$B:=(\beta_\mu, \beta_\sigma, \beta_K, \omega_\mu, \omega_\sigma, \omega_K).$

Our model specification allows for cross-auction heterogeneity in the structural parameters that is observable to potential buyers but unobservable to the econometrician. This is crucial given the considerable variation in assessed price, deal price, area, and location across the auctions in our sample, as shown in \Cref{tab_summary}. 
This also encompasses the possibility that houses sold in different auction rounds are associated with different draws in value distributions or entry costs. For example, given the same observables ($X_i$,$N_i$,$\bm r_i$), a house with a better draw of $\mu_i$ according to \eqref{eq_est} is associated with a more favorable value distribution and may be sold earlier. Conversely, how houses with identical observables in the sample differ in observed entry pattern and transaction time helps us infer the extent of unobserved heterogeneity in value distribution and entry cost parameters.

To estimate our structural model, we adopt the simulated maximum likelihood (SML) method. 
We denote the observable auction outcome for property $i$ by $y_i$, which consists of the number of bidders, the winning bid, and the transaction round. 
Given the outcome $y_i$ and the auction parameters $\Lambda_i$, we can calculate the likelihood of the outcome, denoted by $L_i(y_i|\Lambda_i)$, based on the recurring auction model we discussed in \Cref{sec: model}. The calculation steps are detailed in Online \Cref{subsec:like}.
The log-likelihood function can be written as
\begin{equation}
    L(B;X)=\sum_i^P \log\left[\int L_i(y_i|\Lambda_i)\phi(\Lambda_i|B,X_i)d\Lambda_i\right],
    \label{eq_sml1}
\end{equation}
where $P$ denotes the number of properties and $\phi(\Lambda_i|B,X_i)$ is the probability density of $\Lambda_i$ conditional on $X_i$ and $B$, which can be calculated from \eqref{eq_est}.

Since evaluating the log-likelihood by simulation is computationally expensive,\footnote{To put this into perspective, suppose we draw $S=1,000$ realizations of $\Lambda_i$ to simulate the likelihood for a given $B$. We would need to solve for the equilibrium $S\times P$ times. Given a sample size of 7,772 properties and each solution takes 15 seconds on average, the evaluation would take approximately 1,350 days.} we follow \citet{ackerberg2009new} and \citet{roberts2013should} to reduce the computational burden using importance sampling. Details of the SML method with importance sampling can be found in Online Appendix \ref{Sec: SML}. 

\subsection{Identification}
Although we use a parametric model for estimation, we informally discuss identification here. 
Absent unobserved heterogeneity across auctions, the identification would be straightforward \citep{gentry2014identification}. 
To see the intuition, suppose there is sufficient exogenous variation in equilibrium entry thresholds, which may be driven by variation in the number of potential entrants, entry costs, and reserve prices. 
Then bidders' value distributions can be obtained from empirical bid distributions when the entry threshold (in the first auction round) is sufficiently low. 
As entry thresholds vary, the bid distribution in a particular auction round will be truncated at the corresponding entry threshold. 
Then the expected payoff for the marginal type can be calculated using the recovered value distribution. 
In the final auction round, the expected payoff of the marginal type should equal 0, which helps identify the entry costs. 

Recent empirical studies of auctions have highlighted the importance of accounting for auction-specific heterogeneity that is known to the market participants, but not the econometrician (\citealp{krasnokutskaya2011identification}; \citealp*{athey2011comparing}; \citealp{roberts2013should}). 
In particular, unobserved heterogeneity helps explain within-auction bid correlation conditional on observable sale characteristics. 
Because unobserved heterogeneity renders identification difficult \citep{athey2002identification}, this literature adopts a parametric approach. 
In the home foreclosure auction market, it is plausible that potential buyers may be better informed about some traits of the foreclosed home and the neighborhood than the econometrician. 
We therefore follow the literature and estimate a parameterized model with unobserved heterogeneity. 

The sources of identification for our model parameters are as follows. First, identification of the mean value parameter $\mu$ is based on winning bids or transaction prices. Higher transaction prices correspond to a higher $\mu$. 
Second, the scale parameter $\sigma$ reflects the extent to which potential buyers disagree on the value of a given property, and is identified through variations in transaction prices across auctions, conditional on auction characteristics. 
Third, the entry cost $K$ is identified through potential buyers' entry decisions across auction rounds. If the entry cost is high, potential buyers would be reluctant to enter in the first period and tend to wait until the next period to ensure that there are no strong competitors.

\section{Parameter Estimates and Model Fit}
Having discussed our empirical strategies, we now report our parameter estimates in Table \ref{tab_parameter}.
Estimation results for the recurring auction model are reported in Panel A. As a comparison, we also estimate a single-round model using the same data and the same parametric distribution assumptions. In the single-round model, we ignore the links between auction rounds for the same property and treat them as independent auctions.
Estimation results for the single-round model are presented in Panel B of Table \ref{tab_parameter}.

\begin{table}[ht!]
\begin{center}
    \begin{threeparttable}    
  \centering
  \caption{Parameter Estimates.} \label{tab_parameter}
    \begin{tabular}{lcccccc}
    \toprule
          & constant & $\log\left(\begin{array}{c}\text{assess.} \\
          \text{price}\end{array}\right)$ & log(dist.) & area (100 $m^2$) & $\omega$ & mean \\
          
    \midrule
    \multicolumn{6}{l}{Panel A: Recurring auction model results} \\
    $\mu\sim TRN$    & -0.198 & 0.994 & -0.040 & -0.036 & 0.162 & 4.330  \\
    $ (X\beta_\mu, \omega_\mu, 1, 7) $      & (0.018) & (0.004) & (0.002) & (0.007) & (0.003)  \\
    $\sigma \sim TRN$ & 0.249 & -0.028 & 0.023 & 0.031 & 0.100 & 0.192 \\
    $ (X\beta_\sigma, \omega_\sigma, 0.01, 3)$      &     (0.019) & (0.004) & (0.002) & (0.007) & (0.003) \\
    $K\sim TRN$     &     -2.971 & 0.634 & -0.260 & 0.301 & 0.479 & 0.509 \\
    $ (X\beta_K, \omega_K, 0, 15)$      &     (0.131) & (0.027) & (0.018) & (0.048) & (0.014)\\ [0.5em]
    \multicolumn{6}{l}{Mean of $v\sim TRLN(4.330, 0.192, 10^{-4}, 1200)$: 77.4} \\
    \midrule
    \multicolumn{6}{l}{Panel B: Single-round auction model results} \\
    $\mu\sim TRN $     &      -0.294 & 1.012 & -0.057 & -0.080 & 0.209 & 4.212  \\
    $(X\beta_\mu, \omega_\mu, 1, 7) $      &         (0.019) & (0.004) & (0.003) & (0.008) & (0.003)   \\
    $\sigma \sim TRN $ &         0.190 & -0.018 & 0.020 & 0.037 & 0.083 & 0.180 \\
    $(X\beta_\sigma, \omega_\sigma, 0.01, 3)$      &         (0.014) & (0.003) & (0.003) & (0.007) & (0.002) & \\
    $K\sim TRN$      &         -2.370 & 0.567 & -0.301 & 0.212 & 0.560 & 0.607  \\
    $(X\beta_K, \omega_K, 0, 15)$      &         (0.139) & (0.029) & (0.024) & (0.050) & (0.016) & \\[0.5em]
    \multicolumn{6}{l}{Mean of $v\sim TRLN(4.212, 0.180, 10^{-4}, 1200)$: 68.6}\\
    \bottomrule
    \end{tabular}
\begin{tablenotes} \small
\item[\hspace{-1em}]  Notes: 
(1) Potential buyers' valuation is assumed to follow a truncated lognormal distribution: $v \sim TRLN(\mu, \sigma, 10^{-4},1200)$. (2) Standard errors in parentheses are obtained through bootstrapping 200 times. (3) The rightmost column shows the mean of $\Lambda=\{\mu, \sigma, K\}$ obtained through simulation. (4) All coefficients are significant at the $p<0.01$ level.
\end{tablenotes}
      \end{threeparttable}
\end{center}
\end{table}%

To assess the fit of the two models, we calculate both in-sample and out-of-sample mean squared error of predictions (MSEP) for the deal price and number of bidders.\footnote{MSEP is a commonly used model selection criterion (see, e.g., \citealp{li2009entry}).}
Results are reported in \Cref{table:model_fit}.
Model fit results support the recurring model as the preferred specification, since it outperforms the single-round alternative for predicting both the deal price and number of bidders.


\begin{table}[th!]
\begin{center}
\begin{threeparttable}

\centering
{
\caption{Mean Squared Error of Predictions.}
\label{table:model_fit}
\begin{tabular}{lcclcc}
\toprule
                & \multicolumn{2}{c}{In-Sample} &  & \multicolumn{2}{c}{Out-of-Sample} \\              
                 \cline{2-3} \cline{5-6} 
                                 & Recurring    & Single-round   &  & Recurring      & Single-round     \\ \midrule
Deal price           & 0.0372       & 0.0469         &  & 0.0384         & 0.0482           \\
Number of bidders & 14.857       & 17.060         &  & 15.318         & 17.3161           \\ \bottomrule
\end{tabular}
}
\begin{tablenotes} \small
\item[\hspace{-1em}] Notes: (1) For deal prices, we normalize the error as a percentage of the assessed price. (2) For out-of-sample predictions, we estimate the recurring and the single-round auction model using data from 4,000 randomly selected properties (5,887 auctions). We then compute the MSEP for the 3,772 properties (5,524 auctions) not included in the estimation sample. 

\end{tablenotes}
\end{threeparttable}
\end{center}
\end{table}

As Panel A in \Cref{tab_parameter} shows, the assessed price accurately reflects potential buyers’ mean value parameter $\mu$. The coefficient is close to the hedonic regression results reported in Table \ref{tab_sorting}. This is consistent with the observation that the assessed price strongly correlates with the deal price, which, in turn, reflects the bidders' valuation. 
Conditional on the assessed price, $\mu$ is negatively affected by the property's distance to the city center and its construction area. 
However, the magnitude of the coefficients is small. 
This could mean that the price assessment slightly overvalues the construction area and undervalues the distance to the city center. 

Estimation results for $\sigma$ suggest that there is greater variation in potential buyers' private valuations for properties located in the suburbs and those with larger construction areas. However, there is greater consensus among potential buyers regarding the value of properties with higher assessed prices.

The mean values of $\mu$ and $\sigma$ in our sample are 4.330 and 0.192, respectively. The corresponding value distribution $TRLN(4.330, 0.192, 10^{-4},1200)$ has a mean of 77.4, which implies that the mean of the private value distribution is 774 thousand CNY (114 thousand USD). 

Our estimation indicates that entry costs are higher for properties with higher assessed values, larger construction areas, and shorter distances to the city center. As discussed in \Cref{subsec:indback}, the requirement to deposit 10\% to 20\% of the assessed value and to pay the full amount within 5 working days of winning poses a financial challenge for potential buyers, who must make costly efforts to tackle the liquidity constraint. For properties with higher assessed values, the liquidity constraint binds more tightly. The mean entry cost in our sample is 5,090 CNY (748 USD). 


Panel B in \Cref{tab_parameter} reports the estimation results for a single-round auction model, and reveals the extent of bias when the recurring structure is ignored in auction estimation. The main difference between estimating the recurring auction model and the single-round auction model is that the former is based on observations at the \textit{property} level, whereas the latter is estimated at the \textit{auction} level. As a result, the single-round auction model misses information on the linkage between auction rounds. Since the single-round auction model does not incorporate the sorted entry pattern across auction rounds, it is ``unaware'' of the downward updates in the upper bound of potential buyers' value distribution. By treating subsequent auction rounds in the same way as the initial round, the single-round auction model may well underestimate potential buyers' values and overestimate entry costs. 
As Panel B shows, the mean of $\mu$ is notably lower, and the mean of $K$ is higher than the estimation results for a recurring auction model. In this case, the mean value for a property becomes 686 thousand CNY (101 thousand USD), which is underestimated by 11.4\%. The mean entry cost is 6,070 CNY (867 USD), which represents a 16.1\% overestimation. 

\section{Counterfactuals}
We conduct two counterfactual exercises based on the structural estimation results.
First, to quantify the efficiency and revenue gains associated with recurring auctions, we reduce the number of possible auction rounds and examine the outcomes of single-round auctions and 2-period recurring auctions. 
We keep the reserve prices unchanged for this exercise. Specifically, the reserve prices in the single-round auctions are the observed first-period reserve prices, and the reserve prices in the 2-period recurring auctions are the observed first- and second-period reserve prices. 
Results are presented in Panel A of \Cref{tab_counterfactual}. The rightmost column reports the efficiency and revenue of the observed 3-period recurring auctions. 
Compared with single-round auctions, 3-period recurring auctions increase the efficiency and revenue by 16.6\% and 15.9\%, respectively. This translates to an efficiency gain of 1.46 billion CNY (0.21 billion USD) and a revenue gain of 1.28 billion CNY (0.19 billion USD) from 2017 to 2019 in Fujian province alone.  Extrapolating to the entire country, the annual efficiency and revenue gains amount to 23.13 and 20.19 billion CNY (3.40 and 2.97 billion USD), respectively.\footnote{Note that home foreclosure auctions in Fujian province represent approximately 2.4\% of the total market in China, and home foreclosure auctions in 2019 account for 38\% of the sample. To calculate the annual gain at country level, we multiply the annual gain in Fujian from 2017 to 2019 by 38\% and divide the result by 2.4\%.}

\begin{table}[htbp]
  \begin{center}
      \begin{threeparttable}
        
  \caption{Counterfactual Analyses.} \label{tab_counterfactual}
   
    \begin{tabular}{lccc}
    \toprule
          & Single-round ($T=1$) & Recurring ($T=2$) & Recurring ($T=3$) \\
    \midrule
    \multicolumn{4}{l}{Panel A: current reserve price} \\
    Mean efficiency (10K CNY) & 114.28 & 132.48 & 133.26 \\
    Mean revenue (10K CNY) & 104.06 & 119.98 & 120.62 \\
    \midrule
    \multicolumn{4}{l}{Panel B: optimal reserve price} \\
    Mean efficiency (10K CNY) & 136.88 & 137.68 & 137.72 \\
    Mean revenue (10K CNY) & 123.49 & 124.27 & 124.31 \\
    \bottomrule
    \end{tabular}%
    \begin{tablenotes} 
 \item[\hspace{-1em}]     \small  Notes: (1) We report mean revenue and efficiency at property level. 
 (2) ``Single-round (T=1)'' refers to single-round auctions; ``Recurring (T=2)'' refers to 2-period recurring auctions; ``Recurring (T=3)'' refers to 3-period recurring auctions. 
 (3) For Panel A, we use the current reserve prices, i.e., the reserve prices used in the estimation of the 3-period recurring auctions.
 For Panel B, we use the optimal reserve prices for efficiency and revenue, respectively, in each of the three cases in which $T=1$, $T=2$, and $T=3$.
    \end{tablenotes}
        \end{threeparttable}
  \end{center}
\end{table}%

It is worth pointing out that most of the efficiency and revenue gain from using recurring auctions is realized when there are two possible rounds, i.e., $T=2$. Adding an additional auction round beyond the second improves the auction outcome to a lesser degree. Interestingly, prior to 2017, foreclosed homes were auctioned up to four times in a row. 
Our findings provide justification for the policy change that reduced the number of auction rounds by one.

\begin{figure}[ht!]
    \centering
    \caption{Efficiency-maximizing, Revenue-maximizing, and Observed Reserve Prices (as Fractions of Assessment Price) in 3-Period Recurring Auctions.}\label{fig:res_seq}
    \includegraphics[width = 0.75\textwidth]{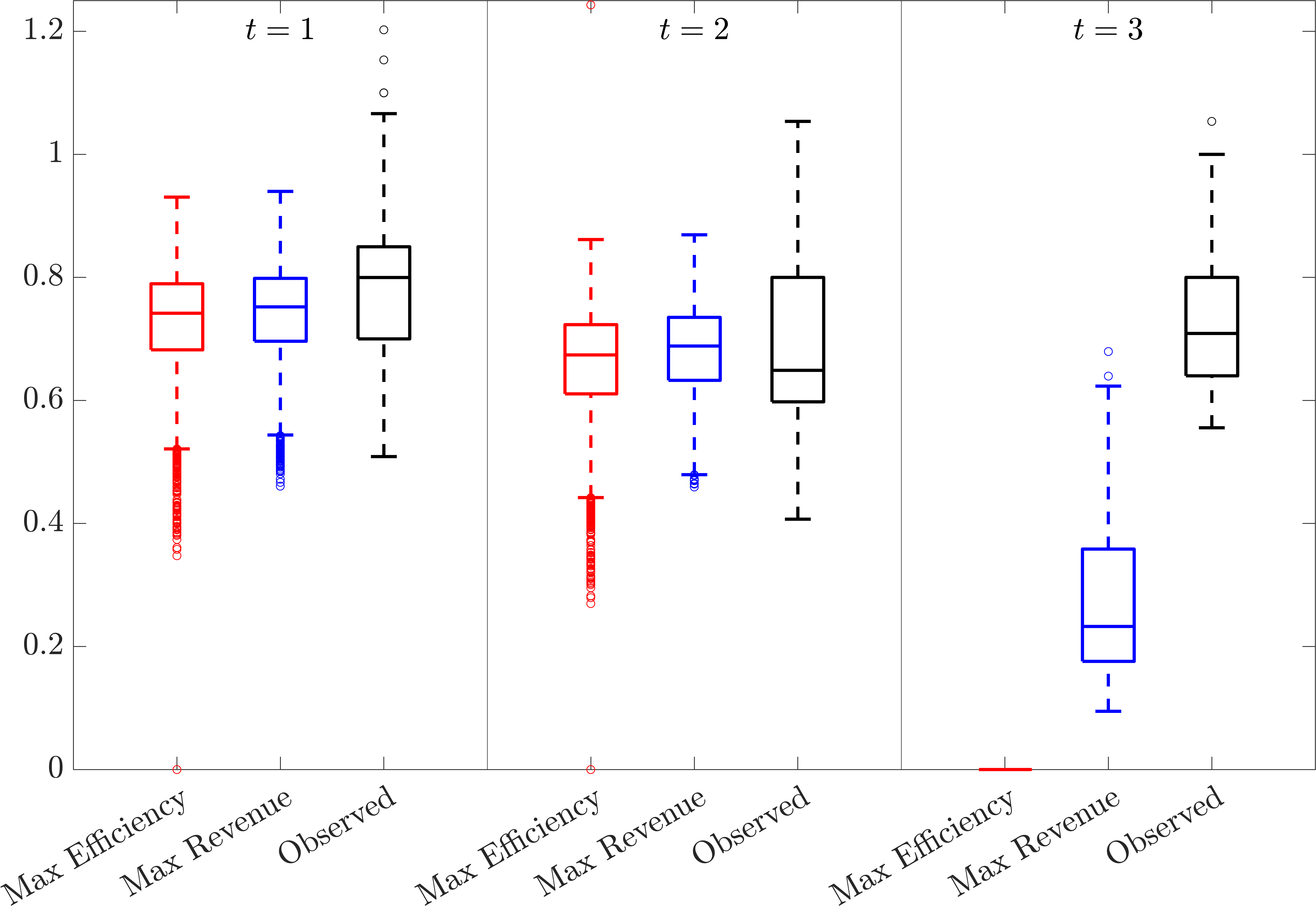}
\end{figure}

Second, we explore optimal auction design in practice by applying the pricing rules in \Cref{thm:eff_design} and \Cref{thm:rev_design}. Results are presented in Panel B of \Cref{tab_counterfactual}. 
\Cref{fig:res_seq} visualizes the reserve prices in 3-period recurring auctions to maximize efficiency or revenue for the properties in our sample. Box plots of the observed reserve prices are also presented. The figure indicates that the observed reserve prices in the first 2 periods are fairly close to optimal for either efficiency or revenue maximization. However, the observed reserve prices in the final period are too high, and cause excessive auction failures and leave room for improvement. 
In fact, optimal reserve price sequences can raise efficiency by 3.35\% and revenue by 3.06\% over actual outcomes. 
On a national scale, these improvements translate to an efficiency gain of 5.43 billion CNY (0.80 billion USD) and a revenue gain of 4.50 billion CNY (0.66 billion USD). 


Recurring auctions continue to outperform single-round auctions in both revenue and efficiency with optimal reserve prices (respectively for both kinds of auctions), but the relative gain is modest.\footnote{To explore how the parameters matter for the efficiency and revenue improvements from adding auction rounds when the reserve prices are optimally set, we regress improvements on the fitted values of $\mu$, $\sigma$, $K$, and number of potential entrants. 
For properties with lower $\mu$ or $\sigma$, or higher $K$ or the number of potential buyers, the extent of efficiency and revenue improvement is lower. Detailed results can be found in Online Appendix \ref{tab_revimprove}. This provides further guidance for the design of recurring auctions in our empirical setting.}
In practice, a seller may still benefit from holding recurring auctions for the following reasons. 
First, setting the reserve price optimally requires detailed information about potential buyers' value distribution for each property. This information may be unavailable in practice, and the fine-tuning of reserve prices may impose additional administration costs. In contrast, as shown in \Cref{tab_counterfactual}, recurring auctions with a rule-of-thumb type of decreasing reserve price sequence perform reasonably well. 
Second, even if the value distribution is known, the seller may still hold recurring auctions to satisfy legal restrictions on reserve prices. 
In our context, the optimal single-round auctions prescribe relatively low reserve prices, and often contradict the mandate that the reserve price in the initial auction must be set above 70\% of the assessed price.\footnote{There may be practical reasons for not running single-round auctions with low reserve prices, such as curtailing corruption as in \citet{cai2013china}.} To put this into perspective, in single-round auctions, the efficiency-maximizing reserve price is 0 and the average revenue-maximizing reserve price is 240 thousand CNY, both of which are nowhere close to the amount required by law, which is 1.005 million CNY. This is less of an issue for recurring auctions, in which the mean efficiency-maximizing reserve price sequence is $(1.057, 0.964, 0)$ in million CNY, and the mean revenue-maximizing reserve price sequence is $(1.074, 0.989, 0.362)$ in million CNY.

We conclude this section with two comments on the effects of entry subsidies and exogenous shocks to entry costs. First, the seller can achieve the same level of efficiency and revenue with optimally designed entry subsidies as using optimally designed reserve prices.\footnote{Negative entry subsidies are allowed and correspond to entry fees.} This is because holding fixed reserve prices, entry subsidies can be used to induce the desired entry thresholds,\footnote{Specifically, the subsidies needed can be solved from \eqref{eqn:rofv} by allowing time-varying entry costs (due to time-varying subsidies) while fixing the reserve price sequence.} and we have shown that the efficiency and revenue are determined by entry thresholds in equilibrium. Second, under observed reserve prices, a 10\% increase in entry costs (509 CNY on average) reduces the average efficiency (revenue) in the $T=1$ auction by 3,906 (3,507) CNY and that in the $T=3$ auction by 1,580 (1,496) CNY.\footnote{A 10\% decrease in entry costs leads to changes in revenue and efficiency of similar magnitudes, albeit in opposite directions. Detailed results are presented in Online Appendix Table \ref{tab_entry}.} This shows that entry costs have non-negligible effects on the auction outcome, but recurring auctions are less affected, reflecting that the sorting pattern in recurring auctions helps potential buyers economize on entry.  


\section{Conclusion}
This paper studies recurring auctions in which an item can be auctioned again if a previous attempt fails.
Our theoretical analysis shows that a recurring auction with an appropriately chosen reserve price sequence can generate greater expected total surplus than any single-round auction. This is because potential buyers sort their entry over time, and participation in later rounds is contingent on no entry in previous rounds. 
Sorted entry generates two benefits for the expected total surplus. First, it allows potential buyers to economize on their entry, since weak potential buyers can wait to enter when the market is less competitive. 
Second, weak potential buyers are encouraged to enter if no one has entered in previous rounds, which reduces the auction failure rate. 
Similarly, recurring auctions can always raise the seller's expected profit compared with single-round auctions. 
In line with our theoretical results, our empirical analysis suggests that recurring auctions lead to large efficiency and revenue gains in China's home foreclosure market. Compared with single-round auctions with the same (first-period) reserve price, recurring auctions raise efficiency by 16.60\% and revenue by 15.92\%. Using the optimal reserve price sequences derived from our model can further improve efficiency by 3.35\% and revenue by 3.06\%, respectively.

\begingroup
\setstretch{1}
\setlength\bibitemsep{0pt}
\bibliographystyle{aernoboldcomma}
\bibliography{reference}
\endgroup

\appendix

\clearpage
\begin{appendices}

\section{Proofs}
\label{app:proofs}
\renewcommand{\thetable}{A\arabic{table}}
\setcounter{table}{0}
\renewcommand{\thefigure}{A\arabic{figure}}
\setcounter{figure}{0}
\setcounter{equation}{0}
\renewcommand{\theequation}{A\arabic{equation}}

\noindent\begin{proof}[Proof of \Cref{prop: eqm}]
It is useful to prove the following lemma first. 
\begin{lemma}
\label{lemma:singlecrossing}
    Suppose $1\leq t<t'\leq T$ or $t=T=t'-1$. Then $\Pi_t(v;\bm v^*)-\Pi_{t'}(v;\bm v^*)$ strictly increases with $v\in[\underline{v}, \overline{v}]$.
\end{lemma}
\begin{proof}
If $t=T=t'-1$, it is easy to see that $\Pi_t(v;\bm v^*)-\Pi_{t'}(v;\bm v^*)=\Pi_T(v;\bm v^*)$  increases with $v$.

     Consider the $1\leq t<t'\leq T$ case. Simple algebra yields that  
\[\frac{d[\Pi_t(v;\bm v^*)-\Pi_{t+1}(v;\bm v^*)]}{dv} = 
\delta^{t-1}\{G[\min\{\max\{v,v_t^*\},v_{t-1}^*\}]-\delta G[\min\{\max\{v,v_{t+1}^*\},v_{t}^*\}]\}
>0.
\]
As a result, $\Pi_t(v;\bm v^*)-\Pi_{t+1}(v;\bm v^*)$ is increasing. Similarly, $\Pi_{t+1}(v;\bm v^*)-\Pi_{t+2}(v;\bm v^*)$, ..., and $\Pi_{t'-1}(v;\bm v^*)-\Pi_{t'}(v;\bm v^*)$ are increasing. 
Since
\[
\Pi_t(v;\bm v^*)-\Pi_{t'}(v;\bm v^*) = \sum_{\tau = t}^{t'-1}
\left[\Pi_{\tau}(v;\bm v^*)-\Pi_{\tau+1}(v;\bm v^*)\right],
\]
$\Pi_t(v;\bm v^*)-\Pi_{t'}(v;\bm v^*)$ is increasing.
\end{proof}

Suppose that all potential buyers other than $n$  use the equilibrium entry and bidding strategy. We verify that the entry and bidding strategy is optimal for potential buyer $n$ by establishing the following three claims.

\begin{claim}
\label{immediate}
    At time $t$, suppose $v_{t-1}^*>v_t^*$. If potential buyer $n$ with $v_n>v_t^*$ had entered, bidding truthfully in the current auction is optimal. 
\end{claim}

\begin{proof}
We prove the claim by induction. Consider the base case of $t=T$. Since there is no continuation play, the game becomes a single-period English auction, so bidding truthfully is optimal for potential buyer $n$.

Suppose the claim holds for all $t'\geq t>1$. We show that it also holds for $t$ in the following.
At time $t$, if other bidders are present in the auction, according to their bidding strategy, the item will be sold in the current period, so there is no continuation play and bidding truthfully is optimal for potential buyer $n$.
If $n$ is the only bidder, he can either bid up to his true value, which entails buying the item at the current reserve price, or let the auction fail. 
In the former case, the payoff is $v_n-r_t$; in the latter case, 
suppose $\tilde t$ is the earliest time after $t$ such that $v_{\tilde t-1}^*>v_{\tilde t}^*$.\footnote{Let $\tilde t=T$ if $v_{t}^*=v_{\tau}^*$ for all $\tau\in\{t+1,\ldots,T\}$.} Then according to the induction hypothesis, potential buyer $n$ will bid and win in some $t''\in\{t+1,\ldots,\tilde t\}$ and the expected payoff is 
$
\delta^{t''-t}\left[
v_n-\int_{v_{t''}^*}^{v_{t''-1}^*}xd\frac{G(x)}{G(v_t^*)}-r_{t''}\frac{G(v_{t''}^*)}{G(v_t^*)}
\right].
$
It suffices to show that 
\begin{equation}\label{eqn:toshow}
    v_n-r_t\geq\delta^{t''-t}\left[
v_n-\int_{v_{t''}^*}^{v_{t''-1}^*}xd\frac{G(x)}{G(v_t^*)}-r_{t''}\frac{G(v_{t''}^*)}{G(v_t^*)}
\right].
\end{equation}

Since $v_{t-1}^*>v_t^*$, by \Cref{def:v}, we have that $\Pi_t(v_t^*;\bm v^*) \geq \Pi_{\tau}(v_t^*;\bm v^*)$ for all $\tau\geq t+1$. This implies that
\[
\begin{split}
    v_t^*-r_t\geq\delta^{t''-t}\left[
v_t^*-\int_{v_{t''}^*}^{v_{t''-1}^*}xd\frac{G(x)}{G(v_t^*)}-r_{t''}\frac{G(v_{t''}^*)}{G(v_t^*)}
\right]
+\frac{K[G(v_{t-1}^*)-\delta^{t''-t}G(v_{t''-1}^*)]}{G(v_t^*)}
\end{split}.
\]
Then \eqref{eqn:toshow} follows from the fact that $v_n>v_t^*$ and $G(v_{t-1}^*)-\delta^{t''-t}G(v_{t''-1}^*)\geq0$.
\end{proof}

\begin{claim}
    Suppose that $v_{t-1}^*\geq v_n>v_t^*$. If an auction is held at time $t$, potential buyer $n$ prefers to enter that auction rather than wait. 
\end{claim}
\begin{proof}
    By \Cref{immediate}, if potential buyer $n$ enters at time $t$, he will bid truthfully and get the expected payoff $\Pi_t(v_n;\bm v^*)/G(v_{t-1}^*)$. 
    
    If $n$ does not enter at time $t$, it is never optimal for him to wait to enter in some future period $t'> t$ and delay bidding further until some $t''>t'$. This follows immediately from \Cref{immediate} for the $v_{t'-1}^*>v_{t'}^*$ case. For the case in which $v_{t'-1}^*=v_{t'}^*$, entering at $t'+1$ and delaying bidding until $t''$ is a better strategy because it allows $n$ to pay entry cost 1 period later. 

    If $n$ waits to enter in some $t'>t$ and bids without delay, the expected payoff is $\Pi_{t'}(v_n;\bm v^*)/G(v_{t-1}^*)$. 
    By \Cref{def:v}, we know that $\Pi_t(v_t^*;\bm v^*) \geq \Pi_{t'}(v_t^*;\bm v^*)$.
    By \Cref{lemma:singlecrossing} and the assumption that $v_n>v_t^*$, we have that $\Pi_t(v_n;\bm v^*)/G(v_{t-1}^*)>\Pi_{t'}(v_n;\bm v^*)/G(v_{t-1}^*)$, i.e., the payoff of entering now is greater than the payoff of waiting to enter in the future. 
\end{proof}

\begin{claim}
    At time $t$, if $v_n\leq v_t^*$, potential buyer $n$ prefers to wait rather than enter the auction. 
\end{claim}
\begin{proof}
    First, consider the situation in which potential buyer $n$ has entered the auction. If other bidders are present, given their strategy, it is optimal for $n$ to bid truthfully and lose. If $n$ is the only bidder, he may choose to buy at the current reserve price or let the auction fail. 
    However, the strategy of entering and letting the auction fail when there is no competition is worse than not entering in the current period. 

    Then it suffices to show that entering to buy when no one else participates is also worse than waiting. The expected payoff from the former is $\Pi_t(v_n;\bm v^*)/G(v_{t-1}^*)$. Because of \Cref{lemma:singlecrossing}, the assumption that $v_n\leq v_t^*$, and the fact that 
    $\Pi_{t}(v_t^*;\bm v^*)\leq\Pi_{t'}(v_t^*;\bm v^*)$ for some $t'>t$,
    we have that $\Pi_t(v_n;\bm v^*)/G(v_{t-1}^*)\leq \Pi_{t'}(v_n;\bm v^*)/G(v_{t-1}^*)$, where the right-hand side is the expected payoff of waiting to bid truthfully at time $t'$. This implies that waiting is better than entering at time $t$. 
\end{proof} 
\end{proof}

\noindent\begin{proof}[Proof of \Cref{thm: eff}]
First, it is useful to note the following result. 
\begin{lemma}[\citealp{samuelson1985competitive}] 
\label{lemma: s}
In a single-round auction, a reserve price equal to the seller's valuation maximizes efficiency.
\end{lemma}
\begin{proof}[Proof of \Cref{lemma: s}] \citet{samuelson1985competitive} provides a proof in the context of procurement auctions. For completeness, a proof for forward auctions is provided here.  

For any reserve price $r\geq0$, in equilibrium, the entry threshold in value $v^*$ satisfies the following indifference condition:
$
(v^*-r)[F(v^*)]^{N-1}=K.
$
That is, potential buyers with values above $v^*$ participate in the auction, while those with values below $v^*$ do not. It is useful to note that at $r=v_s$, the threshold, denoted by $v^{**}$, satisfies
\begin{equation}\label{eqn: os}
(v^{**}-v_s)[F(v^{**})]^{N-1}=K.
\end{equation}

For any given threshold $v^*$, the expected total surplus is 
\[
TS(v^*) = \int_{v^*}^{\overline v}(x-v_s)d[F(x)]^N-N[1-F(v^*)]K+v_s.
\]
Note that 
\[
\frac{\partial TS(v^*)}{\partial v^*} = Nf(v^*)\left\{K - (v^*-v_s)[F(v^*)]^{N-1}\right\}\gtreqless0\iff v^{**}\gtreqless v^*.
\]
Therefore, $TS(v^*)$ is maximized at $v^*=v^{**}$. Since $r=v_s$ induces the threshold $v^{**}$, it maximizes efficiency. 
\end{proof}

By \Cref{lemma: s}, in a single-round auction, efficiency is maximized at the reserve price $r=v_s$. In this case, the entry threshold in value $v^{**}$ is given by \eqref{eqn: os}. The expected total surplus is $TS(v^{**}) = \int_{v^{**}}^{\overline v}(x-v_s)d[F(x)]^N-N[1-F(v^{**})]K+v_s$.

Consider a 2-period recurring auction. If 
$r_1 = v_s$ and $r_2 = v^{**}-K$,\footnote{It is useful to note that by \eqref{eqn: os}, $ v^{**}-K>v_s$.} then the entry threshold in period 1 is exactly $v^{**}$ and no one enters in the second period. That is, $v_1^*=v_2^*=v^{**}$. Therefore, the equilibrium outcome coincides with that in the single-round auction with $r=v_s$. To complete the proof, it suffices to show that lowering $r_2$ improves efficiency. 

Note that the expected total surplus in the 2-period auction is 
\[
\begin{split}
TS(v_1^*,v_2^*) = &\int_{v_1^*}^{\overline v}(x-v_s)d[F(x)]^N+\delta \int_{v_2^*}^{v_1^*}(x-v_s)d[F(x)]^N\\
&-N\left\{[1-F(v_1^*)]K+\delta[F(v_1^*)-F(v_2^*)][F(v_1^*)]^{N-1}K\right\}+v_s.
\end{split}
\]
Simple algebra shows that\footnote{It is useful to note that $\left.\frac{\partial \left\{\int_{v_1^*}^{\overline v}(x-v_s)d[F(x)]^N-N[1-F(v_1^*)]K\right\}}{\partial v_1^*}\right|_{v_1^*=v^{**}}=\left.\frac{\partial TS(v_1^*)}{\partial v_1^*}\right|_{v_1^*=v^{**}}=0$, where the second equality follows from the proof of \Cref{lemma: s}.}
\begin{align*}
\left.\frac{\partial TS(v_1^*,v_2^*)}{\partial v_1^*}\right|_{v_1^*=v_2^*=v^{**}} &= \delta(v^{**}-v_s -K)\frac{\partial [F(v^{**})]^N}{\partial v^{**}},\text{ and }\\
\left.\frac{\partial TS(v_1^*,v_2^*)}{\partial v_2^*}\right|_{v_1^*=v_2^*=v^{**}} &= -\delta(v^{**}-v_s -K)\frac{\partial [F(v^{**})]^N}{\partial v^{**}}.
\end{align*}
Therefore, after a marginal decrease in $r_2$, the change in $TS(v_1^*,v_2^*)$ is 
\[
\delta(v^{**}-v_s -K)\frac{\partial [F(v^{**})]^N}{\partial v^{**}}\frac{\partial (v_1^*-v_2^*)}{\partial r_2}.\footnote{To be precise, the expression $\frac{\partial (v_1^*-v_2^*)}{\partial r_2}$ denotes the left derivative of $(v_1^*-v_2^*)$ with respect to $r_2$. That is, we consider a marginal \emph{decrease} in $r_2$.} 
\]

In fact, given $r_1=0$, a decrease in $r_2$ from $v^{**}-K$ must increase $(v_1^*-v_2^*)$. Otherwise, if $(v_1^*-v_2^*)$ were not increased, no one would enter in the second period. Then the entry threshold in period 1 stays at $v^{**}$, and a potential buyer with value $v^{**}$ gets a payoff of 0. But for the potential buyer, deviating to the second period generates a positive payoff of $\delta [F(v^{**})]^N[v^{**}-(r_2-\Delta r_2)-K] = \delta [F(v^{**})]^N\Delta r_2$. A contradiction.

Since $\frac{\partial (v_1^*-v_2^*)}{\partial r_2}<0$ and $\delta(v^{**}-v_s -K)\frac{\partial [F(v^{**})]^N}{\partial v^{**}}>0$, the change in $TS(v_1^*,v_2^*)$ is positive after a decrease in $r_2$. This completes the proof. 
\end{proof}

\noindent\begin{proof}[Proof of \Cref{thm: rev}]
In \Cref{subsec:design}, we derive the seller's expected profit as a function of $\bm v^*$, \eqref{eqn:rev}. In the $T=1$ case, simple algebra yields that 
\begin{equation}\label{eqn:rev1period}
 \frac{dR(v_1^*;r_1(v_1^*))}{dv_1^*} =  \frac{d[F(v_1^*)]^N}{dv_1^*}\left\{\left[\frac{1}{F(v_1^*)}\right]^{N-1}K-\left[v_1^*-\frac{1-F(v_1^*)}{f(v_1^*)}-v_s\right]\right\}.
\end{equation}
Since $v_1^*=\overline v$ or $v_1^*=\underline v$ does not maximize $R(v_1^*;r_1(v_1^*))$, \eqref{eqn:rev1period} must equal 0 in the optimum. We denote the maximizer by $v^{**}$.

Consider a 2-period recurring auction, with $\bm v^*=(v^{**},v^{**})$. Clearly, this recurring auction generates the same expected profit for the seller as the optimal single-round auction. However, by taking the derivative of \eqref{eqn:rev} with respect to $v_1^*$, we have that 
\[
\frac{dR(\bm v^*;r_1(\bm v^*))}{dv_1^*}\bigg|_{v_1^*=v_2^*=v^{**}} = 
 \delta K\frac{d[F(v^{**})]^N}{dv^{**}}\left\{\left[\frac{1}{F(v^{**})}\right]^{N-1}-1\right\}>0.
\]
As a result, the seller's expected profit can be improved by raising $v_1^*$ in the recurring auction.
\end{proof}

\noindent\begin{proof}[Proof of \Cref{thm:eff_design}]
We prove the theorem in two steps. First, we show that a solution to \eqref{eqn:eff_problem} exists. Second, we establish that \eqref{eqn:eff_foc} is the necessary condition for the optimum and has at most one solution. As a result, the solution to \eqref{eqn:eff_foc} must exist and be optimal. 

\paragraph{Step 1: Existence.} It is useful to note that to maximize efficiency, it is never optimal to have types below $v_s-K$ enter the auction. So it is without loss to impose the constraint $v_T\geq\min\{v_s+K,\underline v\}$. This constraint and the constraints of \eqref{eqn:eff_problem} form a compact set. Since $TS(\bm v^*)$ is continuous in $\bm v^*$, the compactness of the feasible set guarantees the existence of a solution. 

Moreover, the constraint $v_T\geq\min\{v_s-K,\underline v\}$ never binds. That is, $v_T>\min\{v_s-K,\underline v\}$ in the optimum. Suppose to the contrary that $v_T^*=\min\{v_s+K,\underline v\}$ in the optimum. It is without loss to assume that $v_{T-1}^*>\min\{v_s+K,\underline v\}$, because otherwise, if $v_{T-1}^*=v_T^*=\min\{v_s+K,\underline v\}$, we can think of $T-1$ as the last period. For a given $v_{T-1}^*>\min\{v_s+K,\underline v\}$, choosing the last-period entry threshold is like a single-round problem considered in \Cref{lemma: s}. From \eqref{eqn: os}, it follows that the efficiency-maximizing entry threshold should be strictly greater than $\min\{v_s+K,\underline v\}$, contradicting with $v_T^*=\min\{v_s+K,\underline v\}$ being optimal. 

\paragraph{Step 2: The Necessary Condition.} First, we show that the constraint $v_{t-1}^*\geq v_t^*$ does not bind in the optimum. To see that, consider the entry threshold sequence $\bm v^*$ and assume that $v_{t_0}^*=v_{t_0+1}^*$ for some period $t_0$. It suffices to show that $\bm v^*$ is never optimal. 
If $v_{t_0}^*=v_{t_0+1}^*\leq \min\{v_s+K,\underline v\}$, then it follows from Step 1 that $\bm v^*$ cannot be optimal. So we focus on the $v_{t_0}^*=v_{t_0+1}^*> \min\{v_s+K,\underline v\}$ case. If $v_{t_0}^*=v_{t_0+1}^*=v_T^*$, it is like using a single-round auction at time $t_0$. It follows from \Cref{thm: eff} that efficiency can be improved for the subgame starting from $t_0$, so $\bm v^*$ is not optimal. 
If $v_{t_0}^*=v_{t_0+1}^*>v_T^*$, consider an alternative entry threshold sequence in which all entry thresholds after time $t_0$ are moved to 1 period earlier---i.e., $\bm v^{*'}=(v_1^*,\ldots,v_{t_0}^*,v_{t_0+2}^*,\ldots,v_T^*,v_T^*)$. $\bm v^{*'}$ improves over $\bm v^{*}$ since the efficiency gain after time $t_0$ is realized 1 period earlier. 

Since the constraint $v_{t-1}^*\geq v_t^*$ does not bind, the necessary condition is given by the first-order condition of the unconstrained maximization problem. Simple algebra would verify that the first-order condition (with respect to $v_t^*$) is equivalent to \eqref{eqn:eff_foc}. 
\end{proof}

\noindent\begin{proof}[Proof of \Cref{thm:rev_design}]
The proof of \Cref{thm:rev_design} is omitted, since it closely follows that of \Cref{thm:eff_design}.
\end{proof}

\clearpage

\begin{center}
{\LARGE Recurring Auctions with Costly Entry: \\
Theory and Evidence}\\
\bigskip
{\LARGE \bf ONLINE APPENDIX}\\
\bigskip
{\LARGE (Not Intended for Publication)}
\end{center}

In this appendix, we collect the analyses and discussions omitted from the main text.\footnote{%
This note is not self-contained; it is the online appendix of the paper
\textquotedblleft Recurring Auctions with Costly Entry:
Theory and Evidence.\textquotedblright }

\section{Details in Empirical Analysis}
\renewcommand{\thetable}{B\arabic{table}}
\setcounter{table}{0}
\renewcommand{\thefigure}{B\arabic{figure}}
\setcounter{figure}{0}
\setcounter{equation}{0}
\renewcommand{\theequation}{B\arabic{equation}}

%

\subsection{Likelihood Function}
\label{subsec:like}
Four scenarios emerge as the outcomes of recurring auctions. 
The likelihood function for each scenario can be calculated as follows.

\textit{Scenario 1: No one enters for three consecutive auctions.} This implies that no one's valuation is higher than the entry threshold in the last auction ($v_3^*$), so the likelihood is
$L_1=[F(v_3^*)]^N.$

\textit{Scenario 2: Only one potential buyer enters and wins at the reserve price.
} This implies that there is only one potential buyer whose private value is above the entry threshold. The probability for that event is
$L_2={N \choose 1}\ [F(v_{t-1}^\ast)-F(v_t^\ast\ )]F(v_t^*)^{N-1}.$

\textit{Scenario 3: There are multiple entrants, and the deal price is higher than the reserve price.}  We calculate the likelihood as the unconditional probability of $N_e$ entrants, multiplied by the conditional density of the second-highest value being the deal price:
\begin{align*}
    L_3={N \choose N_e} N_e(N_e-1)[F(v_t^*)]^{N-N_e}f(\hat p)[F(\hat p)-F(v_t^*)]^{N_e-2}[F(v_{t-1}^*)-F(\hat p)],
\end{align*}
where $\hat p$ is the observed winning bid.

\textit{Scenario 4:  Zero probability events given the equilibrium}, such as the winning bid being lower than the predicted entry threshold. The likelihood is 0 for these events. However, this does not imply that the simulated likelihood is 0, since this is only for one particular simulation. If, in all simulations for property $i$, there is at least one simulation draw in which the outcome can be rationalized, the simulated likelihood $\frac{1}{S}\sum_s L_{si}$ would be positive.





\subsection{Importance Sampling} \label{Sec: SML}

In this section, we lay out the detailed steps of the simulated maximum likelihood approach with importance sampling. Specifically, we
rewrite the integral in equation (\ref{eq_sml1}) as follows:
\begin{equation}\label{eqn:impsample}
    \int L_i(y_i|\Lambda_i)\phi(\Lambda_i|B,X_i)d\Lambda_i= \int L_i(y_i|\Lambda_i)\frac{\phi(\Lambda_i|B,X_i)}{g(\Lambda_i|X_i)}g(\Lambda_i|X_i)d\Lambda_i,
\end{equation}
where $g(\Lambda_i|X_i)$ is the importance sampling density, which does \textit{not} depend on the parameters $B$. In practice, we pick an initial guess $B_0$ and use $\phi(\Lambda_i|B_0,X_i)$ as the importance sampling density.

We then simulate the right-hand side of \eqref{eqn:impsample} by drawing $S=1,000$ realizations of $\Lambda_i$ according to the importance sampling density, $g(\Lambda_i|X_i)$. 
Compared with $\phi(\Lambda_i|B,X_i)$, the importance sampling density renders $\Lambda_i$ draws independent of $B$.
The simulation is given by  
\begin{equation}\label{eqn:impsample2} 
\frac{1}{S}\sum_s L_{is}(y_{is}|\Lambda_{is}) \frac{\phi(\Lambda_{is}|B,X_i)}{g(\Lambda_i|X_i)},
\end{equation}
where $\Lambda_{is}$ denotes a representative draw. 
The benefit of the importance sampling approach can be clearly seen from \eqref{eqn:impsample2}: When $B$ changes, it is not necessary to draw a new set of realizations of $\Lambda_i$ and reevaluate $L_{is}(y_{is}|\Lambda_{is})$. Instead, the same set of $S=1,000$ simulations can be used and only $\phi(\Lambda_{is}|B,X_i)/g(\Lambda_i|X_i)$ needs to be reevaluated, which is significantly less time-consuming. 

\begin{figure}[ht!]
    \centering
    \caption{Estimation Steps with and without Importance Sampling.}\label{fig_eststeps}
    \subfloat[SML without Importance Sampling.]{
        \includegraphics[width = 0.49\textwidth]{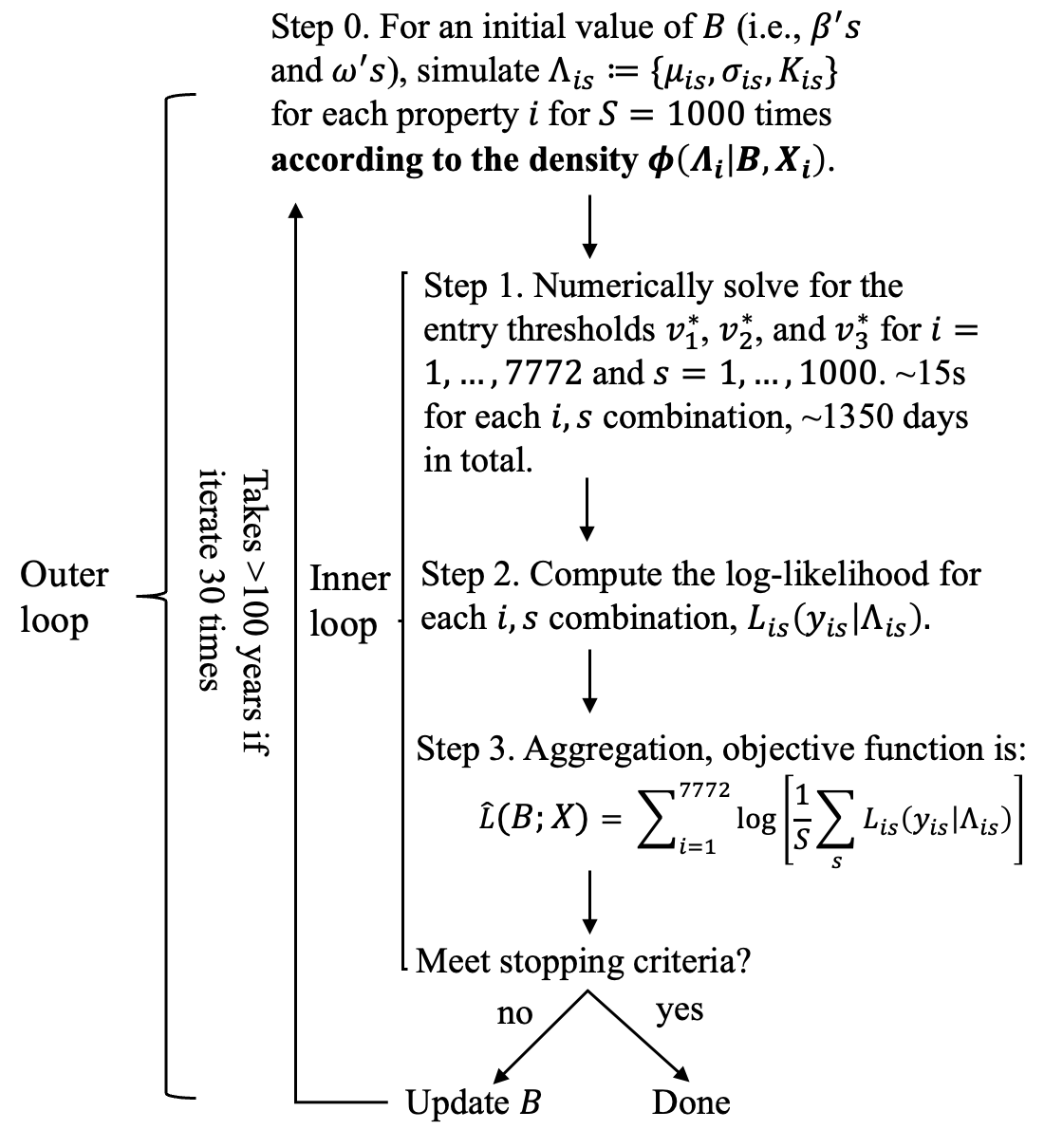}
    }
    \subfloat[SML with Importance Sampling.]{
        \includegraphics[width = 0.49\textwidth]{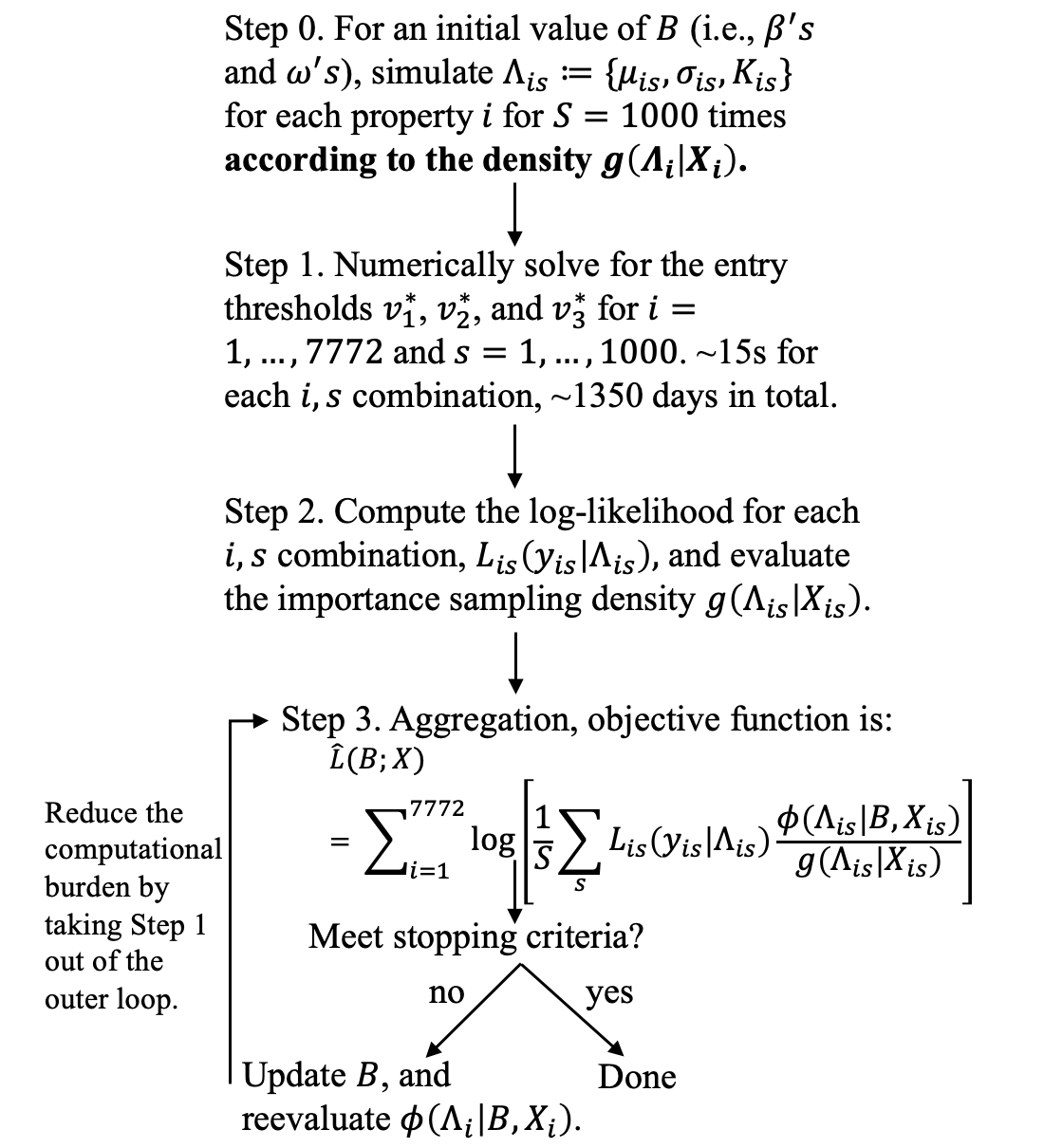}
    }
\end{figure}

\Cref{fig_eststeps} shows the details of the estimation steps. In Panel (a), we show a hypothetical scenario in which we estimate a model without using the importance sampling method. 
The flow chart in Panel (b) of \Cref{fig_eststeps} shows the estimation steps of the simulated maximum likelihood method with importance sampling. 

We use a computer cluster to evaluate $L_{is}(y_{is}|\Lambda_{is})$ for all properties and simulations in parallel, which further reduces the computation time for Step 1 in \Cref{fig_eststeps}(b). After obtaining the results, we search for $B$ that maximizes the simulated likelihood.
Standard errors are computed using a bootstrapping method in which \textit{properties} are resampled 200 times.

\subsection{Alternative Definitions of Potential Buyers}\label{subsec: N robustness}
We examine the robustness of our results to alternative definitions of potential buyers. In the baseline setting, we compute the number of potential buyers as the number of individuals who have shown interest online, divided by 1000, plus the number of actual entrants. In this section, we change the factor from 1/1,000 to 1/500 and 1/1,500 and reestimate the recurring auction model. The estimation results reported in Table \ref{tab_changeN} and the counterfactual analysis results reported in \Cref{tab_counterfactual_altN} are similar to those obtained in the baseline setting, which attests to the robustness of our results. 

\begin{table}[ht!]
\begin{center}
    \begin{threeparttable}    
  \centering
  \caption{Estimation Results for Alternative Definitions of Potential Buyers.} \label{tab_changeN}
    \begin{tabular}{lcccccc}
    \toprule
          & constant & $\log\left(\begin{array}{c}\text{assess.} \\
          \text{price}\end{array}\right)$ & log(dist.) & area (100 $m^2$) & $\omega$ & mean \\
    \midrule
     \multicolumn{7}{l}{Panel A: Factor=1/500} \\
    \multicolumn{1}{l}{$\mu\sim TRN$} & -0.288 & 1.002 & -0.048 & -0.038 & 0.167 & 4.268  \\
      $ (X\beta_\mu, \omega_\mu, 1, 7) $    & (0.022) & (0.005) & (0.003) & (0.010) & (0.003) &  \\
    \multicolumn{1}{l}{$\sigma\sim TRN$} & 0.250 & -0.023 & 0.021 & 0.018 & 0.096 & 0.196  \\
     $ (X\beta_\sigma, \omega_\sigma, 0.01, 3)$     & (0.020) & (0.004) & (0.002) & (0.008) & (0.003)  &  \\
    \multicolumn{1}{l}{$K\sim TRN $} & -2.876 & 0.605 & -0.251 & 0.337 & 0.480 & 0.513 \\
       $(X\beta_K, \omega_K, 0, 15)$   & (0.154) & (0.031) & (0.020) & (0.045) & (0.003)  &  \\
    \midrule
    \multicolumn{7}{l}{Panel B: Factor=1/1500} \\
    \multicolumn{1}{l}{$\mu\sim TRN $} & -0.158 & 0.989 & -0.035 & -0.029 & 0.157 & 4.364  \\
      $ (X\beta_\mu, \omega_\mu, 1, 7) $    & (0.015) & (0.003) & (0.002) & (0.007) & (0.003)  &  \\
    \multicolumn{1}{l}{$\sigma\sim TRN $} & 0.257 & -0.032 & 0.022 & 0.035 & 0.104 & 0.188  \\
      $(X\beta_\sigma, \omega_\sigma, 0.01, 3)$    & (0.018) & (0.004) & (0.002) & (0.008) & (0.004)  &  \\
    \multicolumn{1}{l}{$K\sim TRN$} & -3.011 & 0.642 & -0.249 & 0.304 & 0.478 & 0.514  \\
     $ (X\beta_K, \omega_K, 0, 15)$     & (0.132) & (0.027) & (0.018) & (0.039) & (0.013)  &  \\
    \bottomrule

    \end{tabular}
\begin{tablenotes}
\item[\hspace{-1em}] Notes: (1) We use the same data and the same parameterization for Panel A and Panel B. The only difference lies in the definition of potential buyers. (2) Potential buyers' valuation is assumed to follow a truncated lognormal distribution: $v \sim TRLN(\mu, \sigma, 10^{-4},1200)$. (3) Standard errors in parentheses are obtained through bootstrapping 200 times . (4) The rightmost column shows the mean of $\Lambda=\{\mu, \sigma, K\}$.
\end{tablenotes}
      \end{threeparttable}
\end{center}
\end{table}%

\begin{table}[htbp]
  \begin{center}
      \begin{threeparttable}
        
  \caption{Counterfactual Analyses for Alternative Definitions of Potential Buyers.} \label{tab_counterfactual_altN}
   
    \begin{tabular}{lccc}
    \toprule
          & Single-round ($T=1$) & Recurring ($T=2$) & Recurring ($T=3$) \\
    \midrule
    \multicolumn{4}{l}{Panel A: current reserve price} \\
    \multicolumn{4}{l}{Factor=1/500} \\
    Mean efficiency (10K CNY) & 112.79 & 131.79 & 132.71 \\
    Mean revenue (10K CNY) & 103.37 & 120.30 & 121.09 \\
        \multicolumn{4}{l}{Factor=1/1500} \\
    Mean efficiency (10K CNY) &   114.66 & 132.51 &  133.20\\
    Mean revenue (10K CNY) & 104.13 &  119.55 &  120.11\\
    \midrule
    \multicolumn{4}{l}{Panel B: optimal reserve price} \\
    \multicolumn{4}{l}{Factor=1/500} \\
    Mean efficiency (10K CNY) & 135.74 & 136.70 & 136.75 \\
    Mean revenue (10K CNY) & 123.60 & 124.54 & 124.59 \\
        \multicolumn{4}{l}{Factor=1/1500} \\
    Mean efficiency (10K CNY) & 137.25 &  137.95 & 137.99 \\
    Mean revenue (10K CNY) & 123.16 &  123.85 &  123.89 \\
    \bottomrule
    \end{tabular}%
    \begin{tablenotes} 
 \item[\hspace{-1em}]       Notes: (1) We report the mean revenue and efficiency at property level. 
 (2) ``Single-round (T=1)'' refers to single-round auctions; ``Recurring (T=2)'' refers to 2-period recurring auctions; ``Recurring (T=3)'' refers to 3-period recurring auctions. 
 (3) For Panel A, we use the current reserve prices, i.e., the reserve prices used in the estimation of the 3-period recurring auctions.
A single-round auction is a 3-period recurring auction with the last 2 periods removed. A 2-period recurring auction is a 3-period recurring auction with the last period removed. For Panel B, we use the optimal reserve prices for efficiency and revenue, respectively, in each of the three cases in which $T=1$, $T=2$, and $T=3$.
    \end{tablenotes}
        \end{threeparttable}
  \end{center}
\end{table}%

\subsection{Estimation Results by Year} \label{estimation_year}
In this section, we divide our sample into two groups: houses auctioned in 2017 and houses auctioned in 2018 and 2019. The estimation results reported in \Cref{tab_parameter_year} suggest that coefficient estimates are similar across the two groups. 

\begin{table}[ht!]
\begin{center}
    \begin{threeparttable}    
  \centering
  \caption{Estimation Results by Year.} \label{tab_parameter_year}
    \begin{tabular}{lcccccc}
    \toprule
 & constant & $\log\left(\begin{array}{c}\text{assess.} \\
\text{price}\end{array}\right)$ & log(dist.) & area (100 $m^2$) & $\omega$ & mean \\
\midrule
\multicolumn{7}{l}{Panel A: Year=2017} \\
\multicolumn{1}{l}{$\mu\sim TRN$} &     -0.318 & 1.033 & -0.025 & -0.052 &  0.159 & 4.336  \\ 
$ (X\beta_\mu, \omega_\mu, 1, 7)$ & (0.028) & (0.007) & (0.004) & (0.014)  & (0.005) & \\
\multicolumn{1}{l}{$\sigma\sim TRN $} & 
    0.246 & -0.025 & 0.033 & 0.014 & 0.099 & 0.191  \\
$(X\beta_\sigma, \omega_\sigma, 0.01, 3)$ & (0.032) & (0.007) & (0.004) & (0.014) & (0.006) & \\
\multicolumn{1}{l}{$K\sim TRN $} &     -3.476 & 0.719 & -0.374 & 0.398 & 0.495 & 0.506  \\
$(X\beta_K, \omega_K, 0, 15)$ & (0.142) & (0.033) & (0.036) & (0.064) & (0.022)\\
\midrule
\multicolumn{7}{l}{Panel B: Year=2018 or 2019} \\
\multicolumn{1}{l}{$\mu\sim TRN $} & -0.175 & 0.987 & -0.044 & -0.039 & 0.157 & 4.329 \\
$(X\beta_\mu, \omega_\mu, 1, 7)$ & (0.019) & (0.004) & (0.003) & (0.008) & (0.003)  & \\
\multicolumn{1}{l}{$\sigma\sim TRN $} &   0.257 & -0.031 & 0.020 & 0.038 & 0.099 & 0.192  \\
$ (X\beta_\sigma, \omega_\sigma, 0.01, 3)$ & (0.018) & (0.004) & (0.003) & (0.009) & (0.003)   & \\
\multicolumn{1}{l}{$K\sim TRN $} & -2.890 & 0.618 & -0.235 & 0.294 & 0.477 & 0.514  \\
$(X\beta_K, \omega_K, 0, 15)$  & (0.164) & (0.034) & (0.021) & (0.058) & (0.016)    &  \\
    \bottomrule
    \end{tabular}
\begin{tablenotes}
\item[\hspace{-1em}] Notes: (1) \#Obs.=2,303 in Panel A; \#Obs.=5,469 in Panel B. (2) Potential buyers' valuation is assumed to follow a truncated lognormal distribution: $v \sim TRLN(\mu, \sigma, 10^{-4},1200)$. (3) Standard errors in parentheses are obtained through bootstrapping 200 times. (4) The rightmost column shows the mean of $\Lambda=\{\mu, \sigma, K\}$.
\end{tablenotes}
      \end{threeparttable}
\end{center}
\end{table}%

\subsection{Estimation Results by Time Lag between First Two Auction Rounds} \label{estimation_lag}

To further address concerns about potential buyers or their valuations changing over time, we compare parameter estimates for two subsamples: one with short time lags (bottom quartile, less than 24 days) between the first two auction rounds, and another with long time lags (top quartile, more than 44 days). The results reported in Table \ref{tab_parameter_lag} indicate that the estimated parameters are very similar across subsamples with different time lags, which suggests that new entrants and changing valuations across rounds are not a major concern in the current setting.\footnote{Since the subsamples consist of observations conditional on the failure of the initial auction, estimation results are not expected to be similar to the baseline estimates obtained using the whole sample.}

\begin{table}[ht]
\begin{threeparttable} 
\centering
\caption{Estimation Results by Time Lag.} \label{tab_parameter_lag}
\begin{tabular}{lccccccc}
\toprule
 & constant & log(assess. price) & log(dist.) & area (100 m$^2$) & $\omega$ & mean \\
\midrule
\multicolumn{7}{l}{Panel A: Top quartile (time lag<24 days)} \\
$\mu \sim TRN$ & -0.460 & 1.030 & -0.033 & -0.076 & 0.161 & 4.150 \\
($X_{\beta_\mu}, \omega_{\mu}, 1,7$) & (0.028) & (0.007) & (0.004) & (0.014) & (0.005) & \\
$\sigma \sim TRN$ & 0.276 & -0.023 & 0.028 & -0.006 & 0.096 & 0.199 \\
($X_{\beta_\sigma}, \omega_{\sigma}, 0.01, 3$) & (0.032) & (0.007) & (0.004) & (0.014) & (0.006) & \\
$K \sim TRN$ & -2.042 & 0.434 & -0.201 & 0.415 & 0.478 & 0.591 \\
($X_{\beta_K}, \omega_K, 0,15$) & (0.142) & (0.033) & (0.036) & (0.064) & (0.022) & \\
\midrule
\multicolumn{7}{l}{Panel B: Bottom quartile (time lag>44 days)} \\
$\mu \sim TRN$ & -0.522 & 1.025 & -0.027 & -0.032 & 0.195 & 4.150 \\
($X_{\beta_\mu}, \omega_{\mu}, 1,7$) & (0.019) & (0.007) & (0.004) & (0.014) & (0.005) & \\
$\sigma \sim TRN$ & 0.259 & -0.024 & 0.012 & 0.038 & 0.106 & 0.219 \\
($X_{\beta_\sigma}, \omega_{\sigma}, 0.01, 3$) & (0.032) & (0.007) & (0.004) & (0.014) & (0.006) & \\
$K \sim TRN$ & -1.553 & 0.351 & -0.172 & 0.355 & 0.399 & 0.537 \\
($X_{\beta_K}, \omega_K, 0,15$) & (0.142) & (0.033) & (0.036) & (0.064) & (0.022) & \\
\bottomrule
\end{tabular}

\begin{tablenotes}
\item[\hspace{-1em}] Notes: (1) \#Obs.=617 in Panel A and B; (2) Potential buyers' valuation is assumed to follow a truncated lognormal distribution: $v \sim TRLN(\mu, \sigma, 10^{-4},1200)$. (3) Standard errors in parentheses are obtained through bootstrapping 200 times. (4) The rightmost column shows the mean of $\Lambda=\{\mu, \sigma, K\}$.
\end{tablenotes}

\end{threeparttable}
\end{table}

\subsection{Balance Test for Attrited Houses}
\label{subsec:balance}
A balance test is performed to determine whether the observable characteristics of attrited houses differ significantly from those of other houses. As \Cref{tab_balance} shows, there are no significant differences in any of the observed variables we analyze between attrited houses and others.

\begin{table}[ht!]
\begin{center}
    \begin{threeparttable}    
  \centering
  \caption{Balance Test for Attrited Houses.} \label{tab_balance}
    \begin{tabular}{lccccc}
    \toprule
          & log(reserve) & log(assess. price) & area (100 m$^2$) & log(dist.)  \\
    \midrule
    attrition=1 & 0.010 & -0.002 & -0.013 & 0.001  \\
          & (0.062) & (0.060) & (0.035) & (0.095)  \\
    Year-by-month fixed effects & X     & X     & X     & X      \\
    Observations & 3891  & 3891  & 3891  & 3891  \\
    R-squared & 0.056 & 0.050 & 0.031 & 0.041  \\
    \bottomrule
    \end{tabular}%
\begin{tablenotes}
\item[\hspace{-1em}] Notes: (1) Standard errors in parentheses. (2) We remove houses sold in the first period from the balance test, since no attrition can happen in the first period. 
\end{tablenotes}
      \end{threeparttable}
\end{center}
\end{table}%

\subsection{Determinants of the Number of Bidders}
\Cref{tab_entrants} reports determinants of the number of entrants. The results suggest that the number of potential entrants has a significant and positive impact on the number of actual entrants. The number of potential entrants alone can explain 66\% of the variation in the number of actual entrants (as shown in the first column), which implies that the proxy is well constructed. 

\begin{table}[htbp]
\begin{center}
    
\begin{threeparttable}
    
  \caption{Determinants of the Number of Bidders.} \label{tab_entrants}
    \begin{tabular}{lcccc}
    \toprule
          & \multicolumn{4}{c}{\# of bidders} \\
    \midrule
    \# of potential buyers & 0.538 & 0.605 & 0.621 & 0.593 \\
          &(0.004) &  (0.004) & (0.004) & (0.004) \\
    $\text{round=2} \cdot \text{\# of potential buyers}$ &  &     &       & 0.182 \\
          &  &     &       & (0.009) \\
    $\text{round=3} \cdot \text{\# of potential buyers}$ &   &    &       & -0.143 \\
          &  &     &       & (0.037) \\
    log(assessed price) & & -0.936 & -1.013 & -1.053 \\
          & &(0.029) & (0.029) & (0.029) \\
    area (100 $m^2$)  & &0.379 & 0.510 & 0.534 \\
          & &(0.049) & (0.049) & (0.048) \\
    log (dist to city center) & & -0.131 & -0.141 & -0.144 \\
          & & (0.016) & (0.016) & (0.016) \\
    round=2 & & 1.608 & 1.676 & 0.196 \\
          & &(0.048) & (0.048) & (0.088) \\
    round=3 & & 0.860 & 0.935 & 1.272 \\
          & &(0.081) & (0.080) & (0.165) \\
     Year-by-month fixed effects & & & X &X\\     
    Observations & 11411 & 11411 & 11411 & 11411 \\
    R-squared & 0.659 &0.714 & 0.727 & 0.737 \\
    \bottomrule
    \end{tabular}%
    \begin{tablenotes}
\item[\hspace{-1em}] Notes: (1) Standard errors in parentheses. (2) We pool auctions in all three rounds in this table.  
\end{tablenotes}
\end{threeparttable}
\end{center}

\end{table}%

In the last column, we further explore the heterogeneous effects of the number of potential buyers across auction rounds. The results suggest that each additional potential buyer leads to an increase of 0.593 in the number of actual entrants in the first round. This effect rises to 0.775 (0.593 + 0.182) in the second round and falls to 0.45 (0.593 - 0.143) in the third round. The sorted entry pattern offers an explanation for the differential impact of the number of potential buyers on actual entrants across rounds. Alternatively, if sorting were absent and each auction independently drew a new set of potential buyers, the effect of the number of potential buyers on actual entrants would be homogeneous between the second and third rounds, since the reserve price remains the same.

The increase in the impact of the number of potential buyers on actual bidders during the second round and its decrease in the third round are also consistent with the institutional background and our model. Specifically, there is a 20\% decrease in the reserve price in the second round, with no further decrease in the third round. Therefore, we expect a significant drop in the entry threshold for the second round, while the decrease in the entry threshold between the second and third rounds is more moderate.

\subsection{What Parameters Drive the Efficiency and Revenue Improvements?}

To explore how auction-specific heterogeneity affects revenue and efficiency improvements from adding auction rounds, we regress the percentage improvements in efficiency and revenue—calculated for running $T = 3$ auctions instead of $T = 1$ auction using optimal reserve prices—against the mean value parameter $\mu$, the entry cost $K$, the scale parameter $\sigma$, and the number of potential entrants. Results are reported in Table \ref{tab_revimprove}. 

The findings are consistent with our recurring auction model. Properties with higher mean valuations and greater variance in bidders' value distributions are more likely to be sold, even in a single-round auction. As a result, the revenue and efficiency improvements from the reducing auction failure channel are limited. For properties where potential bidders face higher entry costs or the number of potential entrants is large, adding auction rounds helps potential buyers economize on entry costs and leads to improvements.

         \begin{table}[ht]
          \begin{center}
              \begin{threeparttable} 
  \centering
  \caption{Revenue and Efficiency Improvement by Adding Auction Rounds.} \label{tab_revimprove}
          \begin{tabular}{l*{2}{c}} \toprule
                     &\multicolumn{1}{c}{Eff\_improv}&\multicolumn{1}{c}{Rev\_improv}\\
 \midrule

 $\mu$                &      -0.561&      -0.602\\
                     &     (0.006)   &     (0.006)   \\

 $K$                   &       0.839&       0.898\\
                     &     (0.015)   &     (0.017)   \\
 $\sigma$                  &      -4.282&      -4.421\\
                     &     (0.094)   &     (0.102)   \\
 \# of potential entrants          &       0.027&       0.029\\
                     &     (0.000)   &     (0.000)   \\
 \midrule
 Year-by-month fixed effects & X &X \\
 Observations        &        7699   &        7699   \\
 \(R^{2}\)           &       0.710   &       0.702   \\
  \bottomrule
  \end{tabular}

  \begin{tablenotes}
  \item[\hspace{-1em}] Notes:  (1) Standard errors in parentheses.
  \end{tablenotes}

  \end{threeparttable}
          \end{center}

  \end{table}

\subsection{Varying Entry Costs}

\begin{table}[htbp]
  \begin{center}
      \begin{threeparttable}
        
  \caption{Varying Entry Costs.} \label{tab_entry}
   
    \begin{tabular}{lccc}
    \toprule
          & Single-round ($T=1$) & Recurring ($T=2$) & Recurring ($T=3$) \\
    \midrule
    \multicolumn{4}{l}{Panel A: current entry cost} \\
    Mean efficiency (10K CNY) & 114.28 & 132.48 & 133.26 \\
    Mean revenue (10K CNY) & 104.06 & 119.98 & 120.62 \\
    \midrule
    \multicolumn{4}{l}{Panel B: 90\% of the current entry cost} \\
    Mean efficiency (10K CNY) & 114.67 & 132.70 & 133.42 \\
    Mean revenue (10K CNY) & 104.41 & 120.18 & 120.77 \\
        \midrule
    \multicolumn{4}{l}{Panel C: 110\% of the current entry cost} \\
    Mean efficiency (10K CNY) & 113.89 & 132.26 & 133.10 \\
    Mean revenue (10K CNY) & 103.71 & 119.77 & 120.47 \\
    \bottomrule
    \end{tabular}%
    \begin{tablenotes} 
 \item[\hspace{-1em}]       Notes: (1) We report the mean revenue and efficiency at property level. 
 (2) ``Single-round (T=1)'' refers to single-round auctions; ``Recurring (T=2)'' refers to 2-period recurring auctions; ``Recurring (T=3)'' refers to 3-period recurring auctions. 
 
    \end{tablenotes}
        \end{threeparttable}
  \end{center}
\end{table}%

\end{appendices}

\end{document}